\documentclass[journal]{IEEEtran}
\ifCLASSINFOpdf
\else
\fi

\usepackage{amsmath,amssymb,amsfonts}
\usepackage{amsthm}
\usepackage{algorithmic}
\usepackage{graphicx}

\newtheorem{problem}{Problem} 

\newtheorem{result}{Numerical Result}
\usepackage{algorithm}
\usepackage{algorithmic}
\usepackage{balance}

\usepackage{tikz}
\usepackage{pgfplots}
    \usepgfplotslibrary{groupplots}
\tikzset{every picture/.style={line width=0.75pt}} %set default line width to 0.75pt    
\pgfplotsset{compat=newest}
\usepgfplotslibrary{patchplots}
\usepackage{pgfplotstable}

\pgfplotsset{width=7cm,compat=1.13}
\usepgfplotslibrary{statistics}
  
\usepgfplotslibrary{fillbetween}
\usepackage{filecontents}
\definecolor{blue_c}{RGB}{0, 122, 204}

\newcommand{\reviewed}{black}

\pgfplotstableread[row sep=\\,col sep=&]{
    interval & carT \\

   0.3     & 1   \\
    0.6     & 1  \\    0.9   & 0 \\  
    0.3  & 3 \\
    0.6   & 7   \\    0.9  & 17 \\  1.2 & 33 \\ 1.5 & 84\\ 1.8 & 116 \\ 
    2.1 & 38 \\ 2.4 & 13 \\ 2.7 & 8 \\ 3 & 5 \\
    }\mydata

\begin{document}

\title{A Biologically-Inspired Computational\\ Model of Time Perception}

% Cognitive Modeling of Time Perception Mechanisms  
% A biologically-inspired computational model of time perception mechanisms
% Replication of timing mechanisms\\ and estimation from the behaviour}
% Cognitive modeling time perception
% Cognitive modeling of time perception mechanisms
% Teaching Robots to Perceive Time:\\ A Twofold Learning Approach
% How do you understand time?
% Testing and interpreting time perception in robots
% Replication and estimation of timing mechanisms in robots

\author{In\^es Louren\c{c}o,
		Robert Mattila,
        Rodrigo Ventura,
        and~Bo~Wahlberg% <-this % stops a space
\thanks{This work was partially supported by NewLEADS, the Swedish Research
Council, the Wallenberg AI, Autonomous Systems and Software
Program (WASP), and the FCT project [UID/EEA/50009/2019].}
\thanks{I. Louren\c{c}o, R. Mattila and B. Wahlberg are with the Division of Decision and Control Systems, KTH Royal Institute of Technology, Stockholm, Sweden. E-mails: \{ineslo, rmattila, bo\}@kth.se.}% <-this % stops a space
\thanks{R. Ventura is with the Institute for Systems and Robotics, Instituto Superior T\'{e}cnico, Lisbon, Portugal. E-mail: rodrigo.ventura@isr.tecnico.ulisboa.pt}% <-this % stops a space
}
\maketitle

\begin{abstract}
Time perception -- how humans and animals perceive the passage of time -- forms the basis for important cognitive skills such as decision-making, planning, and communication. 
In this work, we propose a framework for examining the mechanisms responsible for time perception. We first model neural time perception as a combination of two known timing sources: internal neuronal mechanisms and external (environmental) stimuli, and design a decision-making framework to replicate them. We then implement this framework in a simulated robot. 
We measure the agent's success on a temporal discrimination task originally conducted by mice to evaluate its capacity to exploit temporal knowledge.
%To test the ability of the robot to exploit temporal information, we evaluate its performance on a temporal discrimination task originally performed by mice.
We conclude that the agent is able to perceive time similarly to animals when it comes to their intrinsic mechanisms of interpreting time and performing time-aware actions.
Next, by analysing the behaviour of agents equipped with the framework, we propose an estimator to infer characteristics of the timing mechanisms intrinsic to the agents. In particular, we show that from their empirical action probability distribution we are able to estimate parameters used for perceiving time. Overall, our work shows promising results when it comes to drawing conclusions regarding some of the characteristics present in biological timing mechanisms. %, by observing their behavior. %Estimating characteristics of an agent's framework based on its behaviour is one of the goals of current research, which...
%Despite its underlying mechanisms still not being fully understood, they greatly differ from the timing mechanisms inherent to robots. %which helps explain their ineffectiveness in certain perception tasks. 
\end{abstract}
% inherent intrinsic 


% \fi
\IEEEpeerreviewmaketitle

\section{Introduction}
\IEEEPARstart{U}{understanding} different aspects and characteristics of humans and animals has been a driving force in research for centuries (e.g., Skinner's rats \cite{skinner1938behavior}, Pavlov's dogs \cite{pavlov1910work}, or Harlow's monkeys \cite{harlow1965total}). %Edwin Smith Surgical Papyrus, written in the 17th century BC, contains the earliest recorded reference to the brain 
Analysing behaviour can bring insight into how bodies and minds function, from motor impulses to neural mechanisms.  
This can contribute, on the one hand, to obtaining plausible
hypotheses about biological mechanisms and understanding variations to the baseline. Such insights might shed light on, e.g., the causes as well as treatments for diseases. And, on the other hand, replicating these mechanisms in biologically-inspired intelligent agents (e.g., robots) can enhance their cognitive skills and interactions with humans \cite{webb2000does}. Moreover, biologically inspired mathematical models also have the potential to push the boundaries of artificial intelligence. For example, the perceptron \cite{rosenblatt1958perceptron}, which is a mathematical model inspired by the brain's neurons, has been key in the recent advances in deep learning \cite{goodfellow2016deep}.

%Artificially modeling biological features has been extensively discussed in previous works. For example, {\color{red}in 1999,} ...

%These ideas have been present in many previous works: In 199.., and now recently, in 20019, [cite] used the same approach to find that monkeys' actions are based on ... .\\
%- in general, for biology\\
%- for reinforcement learning. e.g., estimate epsilon, delta, etc.\\
%Understanding characteristics about the mechanisms behind human and animal  behaviours has, for a long time, been an important research question. In 199.. [cite] study how many .... 

%Complex systems are represented and simplified using models ....

One of the properties of biological systems whose understanding has seen some advances but is yet not fully understood is \textit{time perception} \cite{soares2016, Gouvea2015}. It concerns the mechanisms responsible for the subjective way time is perceived \cite{allan1979perception}, being in the origin of such idiomatic expressions as ``time flies when you have fun".
%{\color{red}A variable sense of the perception of time , and has been shown to vary as a function of body size and metabolic rate \cite{healy2013metabolic}.} 
Unlike the context-dependent way of perceiving time of biological beings, which is known to be responsible for adaptive behaviours, the basis for the functions of cyber-physical agents such as robots is a perfect linear clock \cite{maniadakis2009explorations}.
%\textbf{On the other hand, cyber-physical agents, such as robots, perform tasks based on state transitions that occur according to linear clock ticks, and lack the ability to perceive time in a context-dependent way \cite{maniadakis2009explorations}.}
Using temporal information in the agents' cognitive processes is considered by many researchers to be one of the milestones in achieving artificial general intelligence \cite{maniadakis2014time, maniadakis2011}. 
When it comes to interacting and blending with humans, temporal information could play an even more important role than spacial information \cite{koene2014relative}. The ability to perceive time would enhance skills like planning, recalling of experiences, and communication, allowing robots to act according to the temporal properties of tasks, adapting to the situation and persons involved and even exploiting the spare time.
 % would enable them to experience the flow of time, perceive synchronization and understand duration. Such capabilities could then enhance their  %Amongst others, time perception is considered to be of utmost importance in collaborative activities and social interactions. 

The first step to incorporate time perception in a robot's decision-making process is to reproduce the way biological systems acquire and use such temporal information \cite{deverett2019interval}.
Many different theories that aim to explain neural timing mechanisms exist, and many cognitive models, such as the internal clock theory \cite{church1984properties} or the behavioural theory of timing \cite{killeen1988behavioral}, have been created.
Furthermore, many computational and robotic models were developed to study the different theories \cite{addyman2016computational}, such as the pacemaker-accumulator \cite{simen2013timescale} and memory decay \cite{addyman2011learning} models. Survey papers like \cite{basgol2021time} have brought these works together and bring us one step closer to understanding what is needed to obtain a complete and explainable framework that exploits temporal information to govern the behaviour of agents.

% The main question we answer in this work is:
In this work, our main goal can be summarized as:
\begin{center}
\textit{Design an explainable end-to-end framework of cognitive mechanisms of time perception whose characteristics can be inferred from the agent's behaviour.}
\end{center}

To answer this question we divide our work in three main parts:
\textit{i)} the design of a framework to model cognitive time perception mechanisms, \textit{ii)} testing the ability of the framework to exploit temporal information, by comparing the behaviour of agents using it to the behaviour of animals performing the same task, and \textit{iii)} estimation of the characteristics of timing mechanisms intrinsic to agents using the framework.

The main contributions provided in each of these parts are, respectively:
\begin{itemize}
	\item A biologically-inspired reinforcement learning framework that replicates neuronal mechanisms of time (the behavior of neurons believed to be responsible for time perception) and combines them with time estimates obtained from the environment. This framework capacitates agents with the ability to perform time-aware actions (Section \ref{sec:create_framework});
	\item Numerical experiments that validate the ability of the proposed framework to exploit temporal information. A simulated robot performing a temporal-discrimination task using the framework demonstrates internal features known to be present in biological timing mechanisms, and estimates the duration of intervals in a similar manner to that of mice on the same task (Section \ref{sec:simulate_framework}).
	\item A method to gain biological insight about the framework used by agents for time perception, based on analysing their behaviour (actions performed) to compute the parameters inherent to their timing frameworks (Section \ref{sec:estimate_framework}). 

\end{itemize}

The two first contributions were initially studied in the conference paper \cite{lourencco2020teaching}, written by the authors, and are further developed in the current paper. The third contribution is completely novel and exploits the previous framework to approach the desired end goal.

 The rest of this paper is structured as follows.
Section \ref{sec:problemform} formulates the time perception problem. Section \ref{sec:create_framework} describes the biologically-inspired decision-making framework proposed to replicate timing mechanisms, and Section \ref{sec:simulate_framework} validates and evaluates the behaviour of an agent using it. Section \ref{sec:estimate_framework} presents the method to estimate parameters of the timing framework from the behavioural analysis. Finally, in Section \ref{sec:conclusions}, the key conclusions of the paper are highlighted and discussed. Moreover, some indications for future work are outlined.

 

%\IEEEpubidadjcol

\section{Problem Formulation}
\label{sec:problemform}

In this section, we start by defining the notation and then present the two main problems addressed in this paper. % together with the experimental setup used to investigate them.

\subsection{Notation}
%All vectors are column vectors, unless transposed. 
The subscript $t$ represents discrete time, and the \textit{i}th element of vector $v_t$ is $v_t(i)$. A general probability mass function is denoted by $p(\cdot)$.
Agents that have time perception are denoted as \textit{timing agents}, and tasks where time perception is needed as \emph{timing tasks}. In the case of interval timing tasks, the variable of interest is the time difference between two events, which we designate $\tau$ and define as the \emph{elapsed time} or \emph{interval duration}.

\subsection{Problem Formulation}
\label{ssec:problem_form}

%Problem 1. Perform actions based on the framwork \\
%The first main challenge we study in this work is the creation of a framework able to exploit temporal information. We meet this aim by replicating and reproducing internal neuronal mechanisms responsible for time perception, in a way that allows agents to perform interval timing tasks.
%Answering the following question originates the complete decision-making framework for the agent to perform actions in these tasks:

The first step of this work is to formalize a framework for modeling the mechanisms responsible for the perception of temporal information. This framework needs to be able to replicate and reproduce the biological mechanisms responsible for time perception, and do so in a way that allows agents to perform interval timing tasks. We formalize this goal in the following question:

%The problem solved by the \textit{TD agent} is formulated as:

\begin{problem}[Biologically-Inspired Timing Framework] How can neural time perception mechanisms be reproduced in a framework that exploits temporal information and produces time-aware behaviours? 
\label{pr:framework}
\end{problem}

By solving Problem \ref{pr:framework}, we establish a decision-making framework that enables agents to perform interval timing tasks based on the replication of neural timing mechanisms, therefore similarly to the way humans and animals perform them.
More specifically, we aim to combine an estimate of the elapsed time obtained from sensory information (external timing) with the agent's biologically-inspired decision-making process (internal timing). The latter includes features that are inspired by the mechanisms believed to govern time perception, such as the dopaminergic activity \cite{soares2016}. As a result, we propose a framework that includes multiple facets of timing mechanisms for performing time-aware actions.

Initial work on this time-perception framework for solving Problem \ref{pr:framework} was presented recently in \cite{lourencco2020teaching} by the authors. In the current paper, we present the framework in Section \ref{sec:create_framework} and provide validating experiments in Section \ref{sec:simulate_framework} that also illustrate some of its key components. 

Once a framework for modeling timing mechanisms exists, the next step is to find an answer for the following question:

\begin{problem}[Estimating Timing Aspects from Behaviour]
How can knowledge about the inner mechanisms of timing agents (such as animals) be gathered from their behaviour?
\label{pr:param_est}
\end{problem}

Finding a solution to Problem \ref{pr:param_est} is the focus of Section \ref{sec:estimate_framework}, where we aim to estimate numerical quantities (that have biological analogues) from observed behaviour of timing agents. In other words, we use our model to estimate information about the intrinsic characteristics of the agents' decision-making process.

In summary, the solution to these two problems results in insight into \emph{i)} how different characteristics of the dopamine system of an agent change its behaviour, as well as, \emph{ii)} which ones are more likely to be the characteristics of the timing mechanisms present in the brain.

%%%%%%%%%%%%%%%%%%%%%%%%%%%%%%%%%%%%%%%%%%%%%%%%%%%%%%%%%%%%%%%%%%%%%%%%%%%%%%%
\section{Biologically-Inspired Timing Framework}
\label{sec:create_framework}

In this section, we review the framework presented in \cite{lourencco2020teaching} to address Problem \ref{pr:framework}.
We first discuss background material and then formally introduce the framework. Subsequently, we explain each of the two main component of the framework, which are the internal and external timing components. The section is concluded with a summary of how they are combined to obtain the complete timing framework.

\subsection{Preliminaries}
\label{ref:framework_summary}
We apply results from neuroscience in a decision-making setup to design a biologically-inspired reinforcement learning \cite{sutton2018reinforcement} algorithm that replicates neuronal timing mechanisms. We denote these mechanisms as \textit{internal timing} mechanisms, since they are related to how internal biological neuronal mechanisms are believed to affect, and enable, the perception of time. In this field, one of the most popular theories is that it is the spiking activity of neurons, mostly known as firing rate, that encodes the passage of time. This is particularly the case for dopaminergic neurons, evidenced by the change in dopaminergic activity when tasks are carried out at different speeds \cite{soares2016}. %These mechanisms allow us to approximately keep track of time even when our sensors are off. 
To replicate mechanisms of time perception we therefore model the principles of dopaminergic behaviour and reproduce them in a simulated robot.

On the other hand, sensory information has been shown to have a direct impact (introduce a bias) on our perception of time \cite{BI}. For example, watching a movie in a faster speed than the natural one leads to an overestimation of time intervals \cite{brown1995time}. We denote the feature responsible for how the perception of time is influenced by external stimuli as \textit{external timing}. 

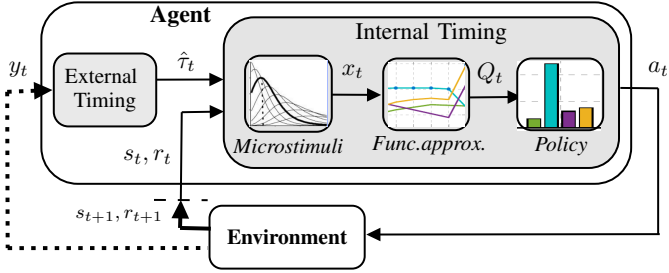
\begin{figure}[t!]
\centering

\tikzset{every picture/.style={line width=0.75pt}} %set default line width to 0.75pt        

\begin{tikzpicture}[x=0.75pt,y=0.75pt,yscale=-0.8,xscale=0.8, every node/.style={scale=0.9}]]
%uncomment if require: \path (0,182); %set diagram left start at 0, and has height of 182

%Rounded Rect [id:dp38118007346288896] 
\draw  [fill={rgb, 255:red, 228; green, 228; blue, 228 }  ,fill opacity=1 ] (313.17,36.47) .. controls (313.17,25.97) and (321.68,17.46) .. (332.17,17.46) -- (543.32,17.46) .. controls (553.82,17.46) and (562.33,25.97) .. (562.33,36.47) -- (562.33,93.49) .. controls (562.33,103.99) and (553.82,112.5) .. (543.32,112.5) -- (332.17,112.5) .. controls (321.68,112.5) and (313.17,103.99) .. (313.17,93.49) -- cycle ;
%Straight Lines [id:da4528546151945758] 
\draw    (381.13,66.25) -- (409.38,66.4) ;
\draw [shift={(412.38,66.42)}, rotate = 180.31] [fill={rgb, 255:red, 0; green, 0; blue, 0 }  ][line width=0.08]  [draw opacity=0] (8.93,-4.29) -- (0,0) -- (8.93,4.29) -- cycle    ;
%Straight Lines [id:da22457804925631986] 
\draw    (468,68) -- (498,67.7) ;
\draw [shift={(501,67.67)}, rotate = 539.4200000000001] [fill={rgb, 255:red, 0; green, 0; blue, 0 }  ][line width=0.08]  [draw opacity=0] (8.93,-4.29) -- (0,0) -- (8.93,4.29) -- cycle    ;
%Rounded Rect [id:dp7065958974260165] 
\draw   (326.58,51.27) .. controls (326.58,45.96) and (330.88,41.67) .. (336.18,41.67) -- (371.65,41.67) .. controls (376.95,41.67) and (381.25,45.96) .. (381.25,51.27) -- (381.25,80.07) .. controls (381.25,85.37) and (376.95,89.67) .. (371.65,89.67) -- (336.18,89.67) .. controls (330.88,89.67) and (326.58,85.37) .. (326.58,80.07) -- cycle ;
%Rounded Rect [id:dp6110633383562631] 
\draw   (413.67,52.47) .. controls (413.67,47.47) and (417.72,43.42) .. (422.72,43.42) -- (458.78,43.42) .. controls (463.78,43.42) and (467.83,47.47) .. (467.83,52.47) -- (467.83,79.62) .. controls (467.83,84.61) and (463.78,88.67) .. (458.78,88.67) -- (422.72,88.67) .. controls (417.72,88.67) and (413.67,84.61) .. (413.67,79.62) -- cycle ;
%Image [id:dp1471971864775039] 
\draw (354.08,65.58) node  {\includegraphics[width=33pt,height=28pt]{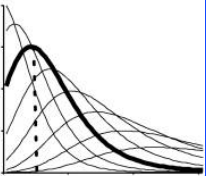}};
%Image [id:dp4019928201100156] 
\draw (440.99,65.96) node  {\includegraphics[width=32pt,height=27pt]{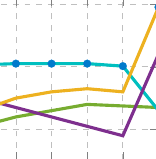}};
%Image [id:dp22577475503589728] 
\draw (524.5,67) node  {\includegraphics[width=35pt,height=29pt]{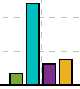}};
%Straight Lines [id:da6757350704014295] 
\draw    (588,154.25) -- (406,154.5) ;
\draw [shift={(403,154.5)}, rotate = 359.91999999999996] [fill={rgb, 255:red, 0; green, 0; blue, 0 }  ][line width=0.08]  [draw opacity=0] (8.93,-4.29) -- (0,0) -- (8.93,4.29) -- cycle    ;
%Rounded Rect [id:dp6262574777634204] 
\draw   (304.17,144.27) .. controls (304.17,139.89) and (307.72,136.33) .. (312.1,136.33) -- (394.07,136.33) .. controls (398.45,136.33) and (402,139.89) .. (402,144.27) -- (402,168.07) .. controls (402,172.45) and (398.45,176) .. (394.07,176) -- (312.1,176) .. controls (307.72,176) and (304.17,172.45) .. (304.17,168.07) -- cycle ;
%Straight Lines [id:da8035508101367428] 
\draw    (563.17,62) -- (587,62.25) ;
%Straight Lines [id:da8962145487486073] 
\draw    (587,62.25) -- (587,154.25) ;
%Straight Lines [id:da8785729148401296] 
\draw [line width=1.5]    (285.58,150.34) -- (305,150.8) ;
%Straight Lines [id:da8201372420607513] 
\draw [line width=1.5]  [dash pattern={on 1.69pt off 2.76pt}]  (176.62,64.46) -- (176.92,163.08) ;
%Straight Lines [id:da4858146073545704] 
\draw [line width=1.5]  [dash pattern={on 1.69pt off 2.76pt}]  (176.62,64.46) -- (201.17,63.49) ;
\draw [shift={(205.17,63.33)}, rotate = 537.74] [fill={rgb, 255:red, 0; green, 0; blue, 0 }  ][line width=0.08]  [draw opacity=0] (11.61,-5.58) -- (0,0) -- (11.61,5.58) -- cycle    ;
%Straight Lines [id:da2321040140262114] 
\draw [line width=1.5]  [dash pattern={on 1.69pt off 2.76pt}]  (176.92,163.08) -- (303.17,163) ;
%Rounded Rect [id:dp9222411714193828] 
\draw   (198.17,30.91) .. controls (198.17,18.27) and (208.42,8.02) .. (221.06,8.02) -- (547.1,8.02) .. controls (559.75,8.02) and (570,18.27) .. (570,30.91) -- (570,99.6) .. controls (570,112.25) and (559.75,122.5) .. (547.1,122.5) -- (221.06,122.5) .. controls (208.42,122.5) and (198.17,112.25) .. (198.17,99.6) -- cycle ;
%Straight Lines [id:da2931173390110142] 
\draw    (261,56.83) -- (310,56.68) ;
\draw [shift={(313,56.67)}, rotate = 539.8299999999999] [fill={rgb, 255:red, 0; green, 0; blue, 0 }  ][line width=0.08]  [draw opacity=0] (8.93,-4.29) -- (0,0) -- (8.93,4.29) -- cycle    ;
%Straight Lines [id:da501341072161938] 
\draw    (286.67,77.32) -- (310.25,76.96) ;
\draw [shift={(313.25,76.92)}, rotate = 539.12] [fill={rgb, 255:red, 0; green, 0; blue, 0 }  ][line width=0.08]  [draw opacity=0] (8.93,-4.29) -- (0,0) -- (8.93,4.29) -- cycle    ;
%Straight Lines [id:da19134449756127392] 
\draw    (286.08,132.75) -- (286.67,77.32) ;
%Straight Lines [id:da18130250217950472] 
\draw [line width=1.5]    (285.58,150.34) -- (285.97,136.75) ;
\draw [shift={(286.08,132.75)}, rotate = 451.63] [fill={rgb, 255:red, 0; green, 0; blue, 0 }  ][line width=0.08]  [draw opacity=0] (11.61,-5.58) -- (0,0) -- (11.61,5.58) -- cycle    ;
%Straight Lines [id:da627847998476788] 
\draw [line width=0.75]  [dash pattern={on 4.5pt off 4.5pt}]  (269.17,132.67) -- (303,132.83) ;
%Rounded Rect [id:dp7962224145377437] 
\draw  [fill={rgb, 255:red, 233; green, 233; blue, 233 }  ,fill opacity=1 ][line width=0.75]  (206,49.49) .. controls (206,44.41) and (210.12,40.29) .. (215.2,40.29) -- (261.38,40.29) .. controls (266.46,40.29) and (270.58,44.41) .. (270.58,49.49) -- (270.58,77.09) .. controls (270.58,82.17) and (266.46,86.29) .. (261.38,86.29) -- (215.2,86.29) .. controls (210.12,86.29) and (206,82.17) .. (206,77.09) -- cycle ;
%Rounded Rect [id:dp921214218248106] 
\draw   (498.17,52.47) .. controls (498.17,47.47) and (502.22,43.42) .. (507.22,43.42) -- (543.29,43.42) .. controls (548.28,43.42) and (552.34,47.47) .. (552.34,52.47) -- (552.34,79.62) .. controls (552.34,84.61) and (548.28,88.67) .. (543.29,88.67) -- (507.22,88.67) .. controls (502.22,88.67) and (498.17,84.61) .. (498.17,79.62) -- cycle ;

% Text Node
\draw (444,27.75) node   [align=left] {Internal Timing};
% Text Node
\draw (353.92,98.44) node  [font=\small] [align=left] {\textit{Microstimuli}};
% Text Node
\draw (442.75,98.17) node  [font=\small] [align=left] {\textit{Func.approx.}};
% Text Node
\draw (386,45.4) node [anchor=north west][inner sep=0.75pt]    {$x_{t}$};
% Text Node
\draw (525.92,98.77) node  [font=\small] [align=left] {\textit{Policy}};
% Text Node
\draw (471,44.4) node [anchor=north west][inner sep=0.75pt]    {$Q_{t}$};
% Text Node
\draw (579,44) node [anchor=north west][inner sep=0.75pt]    {$a_{t}$};
% Text Node
\draw (353.08,156.17) node  [font=\small] [align=left] {\textbf{Environment}};
% Text Node
\draw (248,99.4) node [anchor=north west][inner sep=0.75pt]    {$s_{t} ,r_{t}$};
% Text Node
\draw (286,16.75) node   [align=left] {\textbf{Agent}};
% Text Node
\draw (176,44) node [anchor=north west][inner sep=0.75pt]    {$y_{t}$};
% Text Node
\draw (238.29,63.29) node  [font=\small,color={rgb, 255:red, 255; green, 255; blue, 255 }  ,opacity=1 ] [align=left] {\begin{minipage}[lt]{38.951216pt}\setlength\topsep{0pt}
\begin{center}
\textcolor[rgb]{0,0,0}{External }\\\textcolor[rgb]{0,0,0}{Timing}
\end{center}

\end{minipage}};
% Text Node
\draw (282,37.03) node [anchor=north west][inner sep=0.75pt]    {$\hat{\tau }_t$};
% Text Node
\draw (218,137.4) node [anchor=north west][inner sep=0.75pt]  [font=\footnotesize]  {$s_{t+1} ,r_{t+1}$};

\end{tikzpicture}

\caption{Two-fold time perception framework for replicating time perception mechanisms.
Based on the role of \textit{external environmental stimuli} in time perception, the agent receives environmental observations $y_t$ and estimates the elapsed time $\hat{\tau}_t$. A temporal-difference learning algorithm replicates \textit{internal timing mechanisms} using this estimate. The algorithm uses Microstimuli features \eqref{eq:xfeatures}, which are influenced by the elapsed time estimate $\hat{\tau}_t$ and the state of the environment $s_t$, to compute the Q-values \eqref{eq:Qvalues} of each state-action pair. According to its policy \eqref{eq:actions}, it uses the Q-values to select an action $a_t$ to perform, and receives a reward $r_t$.}
% TE - Time Estimator, MS - Microstimuli, FA - Function Approximation.}
\label{fig:scheme}
\end{figure}

We thus aim for a framework that reproduces the time estimate that stems from internal neuronal processes (studied in Section \ref{ssec:internal}), but also that this estimate should be affected by temporal information collected from the environment (explored in Section \ref{ssec:external}). %Together, they fulfill the ideas that external stimuli influence the perception of time and that context-dependent timing mechanisms can be implemented in non-biological agents.
By combining temporal information that stems from these two different timing sources, we equip robots with their own structure for time perception and show that context-dependent timing mechanisms can be implemented in non-biological agents. A reinforcement learning setup will be used to evaluate the performance of the biologically-inspired decision-making algorithm in timing tasks -- the framework is schematically illustrated in Figure \ref{fig:scheme}, and described in more detail throughout this section.

\subsection{Background}
\label{ssec:background_framework}

Early models of timing in the brain include the clock accumulator, or pacemaker model \cite{killeen1988behavioral}, and the synchronization of brain wave frequencies \cite{matell2004cortico}. However, later models have been created to more precisely match neural responses. 
In \cite{wang2018flexible}, it was found that time is distributively encoded in the dynamics of neural circuits -- the neuronal oscillatory activity, or firing rate.
It was confirmed in \cite{Gouvea2015} that the perception of interval durations can be explained by the speed of change of certain neuronal populations -- particularly in the dopaminergic system, which can therefore be seen as an
internal clock of the brain \cite{Gershman2014}.

Due to its intrinsic signal that represents the disparity between the received and the predicted reward, called a reward prediction error, the dopaminergic system has previously been demonstrated to be involved in reward prediction and action selection \cite{glimcher2011}.
To reproduce its ability to predict the importance of future events from patterns of features that encode the agent's experiences, temporal-difference (TD) learning models have been developed. In such models, the reward prediction error signal is referred to as TD error \cite{sutton1990}.

To represent these features in our framework we use one of the currently most plausible theories when it comes to replicating timing mechansisms, which is called \textit{Microstimuli} \cite{Elliot2008}. Unlike the \textit{Presence} \cite{sutton1990time} and the \textit{Complete Serial Compound} \cite{montague1996, sutton1990} theories, it produces good timing results from a small set of elements per stimulus \cite{Ludvig2012}.
\subsection{Internal Timing}
\label{ssec:internal}

Formally, the TD learning model alluded to in the previous section is modeled in a discrete-time reinforcement learning setting. In such, the interaction between an agent and the environment takes the form of a Markov decision process \cite{puterman2014markov, krishnamurthy2016partially}. 
Formally, the environment is described by a state $s_t \in \mathcal{S}$ that evolves over time, where $\mathcal{S}$ is the state-space and $t$ represents discrete time.
At each time step the agent performs an action, $a_t \in \mathcal{A}$, and receives a reward $r_t$ based on the action performed and the state of the environment. 
Its goal is to find the policy (i.e., a strategy) that maximizes the expected sum of future rewards.
The rest of this section explains the components of the internal timing component shown in Figure~\ref{fig:scheme}.

We use Q-values \cite{sutton2018reinforcement} in our reinforcement learning setup since we are interested in studying the actions performed by the agent. 
To generalise the estimate of the value of a state to states that have similar features, we use \textit{function approximation} \cite{sutton2018reinforcement}, which has been shown very advantageous in problems with large state-action spaces. In particular, we use a linear weighted combination of $D$ features $x(s,a) \in \mathbb{R}^{D}$ to compute the Q-values of state-action pairs, $Q(s,a) : \mathcal{S} \times \mathcal{A} \rightarrow \mathbb{R}$, as
%We consider linear function approximation, where the , is defined as a weighted combination $w$ of the features since in different states and actions some features are valued more than others:
\begin{equation}
Q_t(s,a) = w_t^T x_t(s,a) = \sum_{j=1}^D w_t(j) x_t(j).
\label{eq:Qvalues}
\end{equation}
These features, ${x_t(1),\dots, x_t(D)}$ are chosen so as to replicate the internal neuronal mechanisms mentioned in Section \ref{ssec:background_framework}, which are based on the behaviour of dopaminergic neurons and represented by the \textit{Microstimuli} framework. 

In this framework, each cue and reward deploys a set of $m$ microstimuli. This means that a total of $m \zeta = D$ microstimuli are deployed in episodes with $\zeta$ cues and rewards. 
At time $t$, the level of each microstimulus is represented by a feature $x_t$ (see \cite[Figure 2]{Elliot2008} for a graphical illustration). Mathematically, this relation is modeled as
\begin{equation}
x_{t}(j)= h_{t} \; f\left(h_{t}, \frac{j}{m}, \beta\right), \quad \text{for} \; j=1,\dots,D.
\label{eq:xfeatures}
\end{equation}
In this relation, the level $x_t$ of each microstimulus is computed as the product between an exponentially decaying trace height $h_{t} \in \mathbb{R}$ with decay parameter $\xi$,
\begin{equation}
\centering
\begin{aligned}
h_t &= \exp\{- (1-\xi)t\},
\end{aligned}
\label{eq:features_details1}
\end{equation}
and a Gaussian basis function,
\begin{equation}
\centering
\begin{aligned}
f(h, \nu, \beta) &= \frac{1}{\sqrt{2 \pi}} \exp \left\{ - \frac{(h - \nu)^2}{2 \beta^2}\right\},
\end{aligned}
\label{eq:features_details2}
\end{equation}
with center $\nu$ and width $\beta$.
The amount of decay of each microstimulus can be used as a reference for the elapsed interval due to the steadily decaying memory trace.
As can be seen in \eqref{eq:Qvalues}, each Microstimuli feature from \eqref{eq:xfeatures} is multiplied by a weight $w_t \in \mathbb{R}^D$. These weights, ${w_t (1),..., w_t(D)}$, are the values that the agent wants to learn since they reflect the importance of each feature (such as the strengths of the corticostriatal synapses) for the different state-action pairs~\cite{Elliot2008}.

An essential attribute to have in efficient reward learning is eligibility traces, $e_t$ \cite{singh1996}. Eligibility traces is a characteristic of learning that acts as a vector of memory parameters that are susceptible to changes according to the events they are associated.

In summary, the standard reinforcement learning update equations of the TD error $\delta_t$, the weights $w_t$ and the eligibility traces $e_t$ are, respectively:
\begin{align}
&\delta_t = r_t + \gamma \max_{a} Q(s_{t+1},a)- Q(s_t,a_t), \label{eq:delta}\\
&w_{t+1}(j) = w_t(j) + \alpha \delta_t e_t(j) \label{eq:w}, \\
&e_{t+1}(j) = \gamma \eta e_t(j) + x_t(j),
\label{eq:e}
\end{align}
where $\alpha$ is the learning rate, $\gamma$ is the discount rate, and $\eta$ is a decay parameter that determines the amount of influence of recent stimuli.

Once the Q-values are computed according to \eqref{eq:Qvalues}, the agent is left with the end-goal task of choosing which action to perform. 
We apply a standard action selection mechanism, called $\varepsilon$-greedy \cite{sutton2018reinforcement}. The policy of the agent is to, at each time step, either select a random action or the one with the highest Q-value:
\begin{equation}
a_{t}=\begin{cases}
\arg \underset{a}{\mathrm{\max }} \; Q(s_t,a) , & \text{with probability $1-\varepsilon_{t}$},\\
\text{random action}, & \text{with probability $\varepsilon_{t}$}.
\end{cases}
\label{eq:actions}
\end{equation} 
This corresponds to the exploration-exploitation trade-off that is the basis for learning, where the $\varepsilon_t$ is an exploration parameter with decay parameter $\rho \in [0,1]$ that decreases according to $\varepsilon_{t} = \rho \varepsilon_{t-1}$ as the agent learns.\\

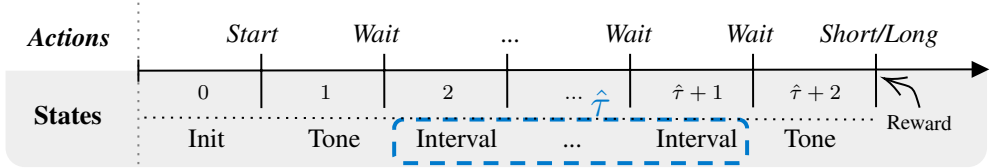
\begin{figure*}[t!]
\centering
%\small

\tikzset{every picture/.style={line width=0.75pt}} %set default line width to 0.75pt        

\begin{tikzpicture}[x=0.75pt,y=0.75pt,yscale=-1,xscale=1]
%uncomment if require: \path (0,89); %set diagram left start at 0, and has height of 89

%Rounded Same Side Corner Rect [id:dp8330154934197067] 
\draw  [draw opacity=0][fill={rgb, 255:red, 238; green, 238; blue, 238 }  ,fill opacity=1 ] (610,74.93) .. controls (610,80.31) and (605.64,84.67) .. (600.27,84.67) -- (124.73,84.67) .. controls (119.36,84.67) and (115,80.31) .. (115,74.93) -- (115,36) .. controls (115,36) and (115,36) .. (115,36) -- (610,36) .. controls (610,36) and (610,36) .. (610,36) -- cycle ;
%Rounded Rect [id:dp5985961770541541] 
\draw  [color=blue_c  ,draw opacity=1 ][dash pattern={on 5.63pt off 4.5pt}][line width=1.5]  (311,65.07) .. controls (311,62.64) and (312.97,60.67) .. (315.4,60.67) -- (484.6,60.67) .. controls (487.03,60.67) and (489,62.64) .. (489,65.07) -- (489,78.27) .. controls (489,80.7) and (487.03,82.67) .. (484.6,82.67) -- (315.4,82.67) .. controls (312.97,82.67) and (311,80.7) .. (311,78.27) -- cycle ;
%Straight Lines [id:da7146759280445087] 
\draw [color={rgb, 255:red, 128; green, 128; blue, 128 }  ,draw opacity=1 ] [dash pattern={on 0.84pt off 2.51pt}]  (182,1.67) -- (182,84.67) ;
%Straight Lines [id:da28358027591939106] 
\draw    (182,36) -- (310.17,35.8) -- (608.17,35.34) (243.97,16.9) -- (244.03,54.9)(305.97,16.81) -- (306.03,54.81)(367.97,16.71) -- (368.03,54.71)(429.97,16.61) -- (430.03,54.61)(491.97,16.52) -- (492.03,54.52)(553.97,16.42) -- (554.03,54.42) ;
\draw [shift={(611.17,35.33)}, rotate = 539.9100000000001] [fill={rgb, 255:red, 0; green, 0; blue, 0 }  ][line width=0.08]  [draw opacity=0] (8.93,-4.29) -- (0,0) -- (8.93,4.29) -- cycle    ;
\draw [shift={(182,36)}, rotate = 539.9100000000001] [color={rgb, 255:red, 0; green, 0; blue, 0 }  ][line width=0.75]    (0,5.59) -- (0,-5.59)   ;
%Curve Lines [id:da5258176305463109] 
\draw    (576,54.67) .. controls (572.16,41.23) and (572.93,45.45) .. (559.72,40.5) ;
\draw [shift={(558,39.83)}, rotate = 381.8] [color={rgb, 255:red, 0; green, 0; blue, 0 }  ][line width=0.75]    (10.93,-4.9) .. controls (6.95,-2.3) and (3.31,-0.67) .. (0,0) .. controls (3.31,0.67) and (6.95,2.3) .. (10.93,4.9)   ;
%Shape: Rectangle [id:dp886502478070782] 
\draw  [color={rgb, 255:red, 255; green, 255; blue, 255 }  ,draw opacity=1 ][fill={rgb, 255:red, 255; green, 255; blue, 255 }  ,fill opacity=1 ] (199,5.67) -- (603,5.67) -- (603,27) -- (199,27) -- cycle ;
%Straight Lines [id:da7626842785734733] 
\draw  [dash pattern={on 0.84pt off 2.51pt}]  (185,59.67) -- (554,60.67) ;

% Text Node
\draw (125,13) node [anchor=north west][inner sep=0.75pt]   [align=left] {\textbf{\textit{Actions}}};
% Text Node
\draw (128,52) node [anchor=north west][inner sep=0.75pt]   [align=left] {\textbf{States}};
% Text Node
\draw (225,11) node [anchor=north west][inner sep=0.75pt]   [align=left] {\textit{Start \ \ \ \ \ \ \ Wait \ \ \ \ \ \ \ \ \ \ ... \ \ \ \ \ \ \ \ Wait \ \ \ \ \ \ \ Wait \ \ \ \ Short/Long}};
% Text Node
\draw (201,64) node [anchor=north west][inner sep=0.75pt]   [align=left] {\ Init \ \ \ \ \ \ \ \ Tone \ \ \ \ \ Interval \ \ \ \ \ \ ... \ \ \ \ \ \ \ Interval \ \ \ \ Tone };
% Text Node
\draw (203,41.4) node [anchor=north west][inner sep=0.75pt]  [font=\footnotesize]  {$\ \ 0\ \ \ \ \ \ \ \ \ \ \ \ \ \ \ 1\ \ \ \ \ \ \ \ \ \ \ \ \ \ \ 2\ \ \ \ \ \ \ \ \ \ \ \ \ \ \ ...\ \ \ \ \ \ \ \ \ \ \ \ \hat{\tau } +1\ \ \ \ \ \ \ \ \ \hat{\tau } +2\ \ \ \ $};
% Text Node
\draw (558.12,56.62) node [anchor=north west][inner sep=0.75pt]  [font=\footnotesize,rotate=-1.01] [align=left] {Reward};
% Text Node
\draw (408.5,44) node [anchor=north west][inner sep=0.75pt]  [font=\Large,color=blue_c  ,opacity=1 ]  {\textbf{$\hat{\tau}$}};

\end{tikzpicture}

\caption[RL task]{Optimal sequence of actions (top row) performed during an episode, and corresponding state transition (bottom row). Once the \textit{Start} button is pressed, the environment changes to the \textit{Tone} state. From there the agent must perform the ``Wait" action until reaching the second ``Tone" state, which happens during a number of \textit{Interval} states sampled is uniformly from the maximum interval length. % and denoted as $I$, $I \in \mathbb{R}$.
After the second \textit{Tone} state, the agent then performs the \textit{Short} or \textit{Long} action that corresponds to its estimated number of \textit{Interval} time steps, $\hat{\tau}$. If the correct action is chosen, the agent receives a positive reward.}
\label{fig:RLtask}
\end{figure*}

This concludes our decision-making framework for replicating internal neuronal mechanisms, {\color{\reviewed}which can be classified as an \textit{intrinsic} and \textit{emergent} timing framework \cite{basgol2021time}}. %Below, we extend this framework in a way that makes it possible for the time estimates to be influenced by environmental stimuli. %For that, in the next section we explain how time estimates can be computed from environmental stimuli.

\subsection{External Timing}
\label{ssec:external}

We now briefly outline how an estimate of the elapsed time can be computed based on environmental stimuli (full details are available in Appendix \ref{sec:appendix_external}).
The complete framework we propose then consist in merging the two estimates (internal and external) to form a biologically-inspired perception of the elapsed time. 

%To compute an estimate of the elapsed time, $\hat{\tau}$, from raw information collected by the agent's sensors, we pose the question: given a set of observations, $\mathcal{O}$, collected during a certain interval of length $\tau$, estimate the interval duration.
It has been shown that a sensory estimate of the passage of time can be obtained by environmental observations $\mathcal{O}$ \cite{Eagleman2004, Brown1995}.
{\color{\reviewed}We assume an innate external timing mechanism that uses} a probabilistic expectation of stimulus change in the environment to compute an estimate of the elapsed time as in \cite{BI}, where a Bayesian framework was proposed to estimate the elapsed time $\tau$ from the environmental data $\mathcal{O}$.

Essentially, we model the joint distribution of the sensory data as Gaussian processes with an Ornstein-Uhlenbeck kernel function. We compute the hyperparameters of the model using Bayesian model selection, and from the maximum likelihood principle obtain the estimate $\hat{\tau}$.

A more detailed explanation about how the elapsed time estimate can be computed was presented in \cite{lourencco2020teaching}, and a summary can, for completeness, be found in Appendix \ref{sec:appendix_external}.
%From \eqref{eq:Gderivates} we estimated the maximum likelihood model parameters $\lambda$ and $\sigma$ of the collected data and used these to estimate the elapsed time $\hat{\tau}$. 

%For example, the kernel matrix expresses how much the process changes from one time step to the next, corresponding to the rhythm of change of the natural environment.
%If the statistical properties remain constant in time, the process is stationary and thus the observations can be modelled as stationary Gaussian processes with an OU kernel \cite{gaussian2007}. 

\subsection{Summary of the Complete Timing Framework}

An agent is navigating the environment and receiving stimuli. Using the external timing mechanism, it is able to estimate the perceived elapsed interval length $\hat{\tau}$ between each stimuli.
Based on the internal timing mechanisms, sets of $m$ decaying microstimuli features, $x_t$ from \eqref{eq:xfeatures}, are deployed when the agent estimates having received the stimuli (cues and rewards). The perceived interval $\hat{\tau}$ then conditions the microstimuli features $x_t$ that replicate the dopaminergic activity believed to regulate timing mechanisms, which influence the $Q$-values of the state-action pairs according to \eqref{eq:Qvalues} and, as can be seen in \eqref{eq:actions}, also the action $a_t$ selected by the agent. 
In summary, this results in a decision-making framework that incorporates external and internal timing mechanisms -- {\color{\reviewed}where the former is responsible for adjusting our perception to the world that surrounds us and the latter for learning time as a result of neuronal changes even when, for example, we are sleeping.}

We assess the complete timing framework in the next section, by analysing the behaviour of the agent when performing a temporal discrimination task.

\section{Simulating the Time Perception Framework }
\label{sec:simulate_framework}

As in \cite{lourencco2020teaching}, we {\color{\reviewed}simulate} a robot that following the framework presented in Section \ref{sec:create_framework} performs sequences of actions in a temporal discrimination task {\color{\reviewed}of duration comparison using prospective timing.}
%
%The actions depend on the state sequence seen by the robot, which is influenced by the internal estimate of time between stimuli.
We start by presenting the experimental setup used, then details on how the framework is simulated, and finally the numerical results that highlight the success of the robot in the task.
The success is measured by the similarity of its actions to the actions performed by a biological timing agent (an agent with temporal cognition, which, in this case, is a mouse) in the same task.

\begin{algorithm}[t!]
%\small
	\caption{Temporal discrimination task}
	\label{alg:RL}
	\begin{algorithmic}[1]
		%		\Require
		\STATE Initialize $Q(s,a) = 0$, for all $s \in \mathcal{S}$, $a \in \mathcal{A}$, and $w(1),\dots,w(D)$ randomly (e.g., $w(j) \in [0,1]$)
		\FOR{each episode}
			\STATE Initialize the state to ``Init"
		\FOR{each time step $t$}
					\STATE Update $x_t(1), \dots ,x_t(D)$ according to \eqref{eq:xfeatures}
			\IF{only one ``Tone" state has passed}
				%\STATE (action should be random walk)
			%\ELSIF{has received stimulus 1 but not 2}
				\STATE Collect data $y_t(1), \dots ,y_t(M)$
				%\STATE (action should be press button short/long)
			\ELSIF{both ``Tone" states have passed}
				\STATE Estimate the elapsed time, $\hat{\tau}$, by maximizing \eqref{eq:equationBI}
				\STATE Update $x_{\hat{\tau}}(1), \dots ,x_{\hat{\tau}}(D)$, according to \eqref{eq:xfeatures}
				%\STATE (action should be press button short/long)		
			\ENDIF
			\STATE Compute the Q-values according to \eqref{eq:Qvalues} and choose $a_t$ according to \eqref{eq:actions}. Take action $a_t$, observe $r_t$, $s_{t+1}$
			\STATE $\delta_t \gets r_t + \gamma \max_{a} Q(s_{t+1},a)- Q(s_t,a_t)$
			\STATE $w_{t+1}(j) \gets w_{t}(j) + \alpha \delta_t e_{t}(j)$, \quad for $j=1,\dots,D$
			\STATE $e_{t+1}(j) \gets \gamma \eta e_{t}(j) + x_{t}(j)$, \;\;\;\quad for $j=1,\dots,D$
			\STATE $s_t \gets s_{t+1}$
		\ENDFOR
		\STATE Until $s_t$ is terminal
		\ENDFOR
	\end{algorithmic}
\end{algorithm}

\subsection{Experimental Setup}
\label{ssec:experimental}

We evaluate the framework proposed in Section \ref{sec:create_framework} in a temporal discrimination setup, where the ability of {\color{\reviewed}an agent} to distinguish time intervals (durations) is evaluated and compared to that of mice performing the same task \cite{soares2016}.
 
In the original experiment \cite{soares2016} three buttons are available to a mouse: ``Start", ``Short" and ``Long". The experiment starts when the animal presses the former, and two auditory tones, separated by a certain interval that varies from episode to episode, are played to the mouse. 
It then has the possibility to press the button that corresponds to its estimated interval length between both tones (``Short" or ``Long").
A reward of water or food is given to the animal if the correct button is pressed (i.e., if the animal correctly estimated the elapsed time between the two tones). 

We use the same setup for our simulated robot.  %Using a temporal-difference learning algorithm, the robot should use its estimate of the elapsed time to reach a certain goal. 
At each time step, the environment can be in one of the states $\mathcal{S}=$\{Init, Tone, Interval\}, and the robot can perform one of the actions $\mathcal{A}=$\{Start, Wait, Short, Long\}. 
Figure \ref{fig:RLtask} schematically shows the optimal sequence of actions in an episode.
The number of ``Interval" states is what defines the \textit{interval duration} $\tau$ of the episode. If we define $L\in \mathbb{N}$ as the experiment's maximum interval duration between tones, then in each episode, $\tau$ is a realization of a discrete uniform random variable $\tau \sim \text{unif}\{1,L\}$.
 %The classification is therefore given by 
%Intervals shorter than L/2 are ``Short", and intervals between L/2 and L are ``Long", being L/2 the classification boundary.
The temporal discrimination task consists of classifying the interval duration as
\begin{equation}
\begin{cases}
\text{``Short"} , & \text{if} \; \tau \in [1,2,\dots, \lfloor \frac{L}{2} \rfloor],\\
\text{``Long"}, & \text{if} \; \tau \in [\lceil \frac{L}{2}+1 \rceil, \dots, L],
\end{cases}
\label{eq:shortvslong}
\end{equation} 
%$\{\text{Short}:=\Delta_t \in [1,2,\dots,L/2]\}$ 
where $\{\lfloor \frac{L}{2} \rfloor, \lceil \frac{L}{2}+1 \rceil\}$ is the classification boundary for $L$ even and $\lceil \frac{L}{2} \rceil$ for $L$ odd.
%
%The choice of the interval duration is done based on 50\% of probability of being short, and 50\% of being long. Inside the chosen interval, there is a uniform probability of choosing any value within that interval. 
In the numerical experiments below, we selected the maximum interval duration $L$ as in the real experiment \cite{soares2016}.
This means that $L=3$ seconds, or equivalently, $L=8$ time steps, corresponding to ``Short" $:=\tau \in \{1,2,3,4\}$ and ``Long" $:=\tau \in \{5,6,7,8\}$.

The increased complexity of the problem comes from the fact that before learning to distinguish between short and long intervals, the robot first has to compute its estimate of the elapsed time between the two stimuli, $\hat{\tau}$, using the external timing mechanism from Section \ref{ssec:external}. Only then can it use the estimate to learn the classification task. The robot computes its estimate of the elapsed time between stimuli by navigating around the environment and gathering data from its sensors during the interval $\tau$ between which the two stimuli are presented to it.
 %This is addressed in Result \ref{result:1} next.
%The choice of the interval duration is done based on 50\% of probability of being short, and 50\% of being long. Inside the chosen interval, there is a uniform probability of choosing any value within that interval. 
%, where the black line represents the passage of time. 

%Eligibility traces, denoted by $e$, are an essential feature of reward learning that, when multiplied by the error, expand the influence of the presence of a state through time \citep{BI}. We consider the update rule:
%\begin{equation}
%e_{t}(s,a)=\begin{cases}
%1+ \gamma \Lambda e_{t-1}(s,a) , & \text{if $s=s_{t}, a= a_{t}, Q_{t-1}(s_{t},a_{t}) = \underset{a}{\mathrm{\max }} Q_{t-1}(s_{t}, a)$ }\\
%0, & \text{if $Q_{t-1}(s_{t},a_{t}) \neq \underset{a}{\mathrm{\max }} Q_{t-1}(s_{t}, a)$} \\
%\gamma \Lambda e_{t-1}(s,a), & \text{otherwise}
%\end{cases},
%\label{eq:etraces}
%\end{equation}
%where $\gamma$ is the discount rate and $\Lambda$ the decay parameter that determines the plasticity window of recent stimuli.

%The action selection mechanism considered is the $\epsilon$-greedy, given by
%\begin{equation}
%a_{t}=\begin{cases}
%\arg \underset{a}{\mathrm{\max } } Q(s_t,a) , & \text{with probability $1-\epsilon_{t}$}\\
%\text{random action}, & \text{with probability $\epsilon_{t}$}
%\end{cases},
%\label{eq:action}
%\end{equation}
%where $\epsilon_t$ decays according to $\epsilon_{t} = \Gamma \epsilon_{t-1}$ and $\Gamma	$ is the decay parameter.
%

\begin{figure}[t!]
\centering
\begin{tikzpicture}
    \begin{groupplot}[
        group style={
            % set how the plots should be organized
            group size=1 by 3,
            % only show ticklabels and axis labels on the bottom
            x descriptions at=edge bottom,
            % set the `vertical sep' to zero
            vertical sep=0pt,
        },
        % set the x axis default to logarithmic
        grid,
        height=2.7cm,
       	width=8cm,
       	ymax=1,
       	xmax=5,
       	xmin=0,
       	ytick={0,0.5},
        xtick ={0, 1, 2, 3, 4},
        xlabel={Time steps, $t$},
        grid style=dashed,
        %legend pos=north west,
        %legend style={draw=none, font=\footnotesize},
    ]
    % start the first plot
    \nextgroupplot[
    ]
        \addplot [color=blue_c, ultra thick,]  coordinates { (0,0)(1,0)(2,0)(3,0)(4,0)(5,1) }; %\addlegendentry{Episode 15} %\footnotesize
        \node[below right, align=center, text=black]
at (rel axis cs:0.05,0.9) {\text{Episode 15} };
    % start the second plot
    \nextgroupplot[ ylabel={TD error, $\delta_t$}
    ]
        \addplot [color=blue_c, ultra thick] coordinates { (0,0)(1,0)(2,0)(3,0)(4,0.05)(5,0.15) }; %\addlegendentry{Episode 250} %\footnotesize
        \node[below right, align=center, text=black]
at (rel axis cs:0.05,0.9) {\text{Episode 250} };
    \nextgroupplot[
    ]
        \addplot [color=blue_c, ultra thick] coordinates { (0,0)(1,0)(2,0)(3,0)(4,0.1)(5,0) }; %\addlegendentry{Episode 400}  %\footnotesize
        \node[below right, align=center, text=black]
at (rel axis cs:0.05,0.9) {\text{Episode 400} };
    \end{groupplot}
\end{tikzpicture}
\caption{Evolution of the TD error throughout three episodes with $\tau = 2$ time steps in different phases of learning. Unsurprisingly, the more the agent has learned the more it expects a reward after correctly classifying the interval ($t=4$), leading to a decreasing TD error at the end of the episode and an increasing one upon receiving the second tone.} %The second tone happens in time step four and the reward is given in five.}
\label{fig:TDerror1}
\end{figure}
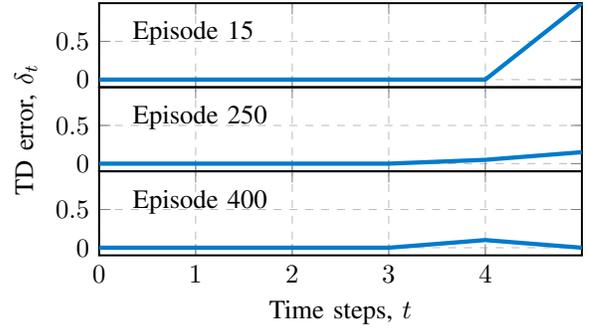

\subsection{Simulating the Proposed Framework}
\label{ssec:simulating}

Following this setup, in the external timing mechanism we chose the values of
\textit{i} LIDAR angles of the simulated robot (emulating the visual system of a mouse) to be the environmental data $y_t(i)$ at time $t$. % which are collected while the robot does the ``Wait" action between tones.
We estimate the elapsed time $\hat{\tau}$ using the maximum likelihood parameters computed for the model of sensory data (see Appendix \ref{sec:appendix_external}).	
As a result of its internal timing mechanisms, sets of $m = 8$ microstimuli features $x_t$ are deployed when the agent receives each of the two stimuli, separated by $\hat{\tau}$ time steps. %The features influence the Q-value of the state-action pair according to \eqref{eq:Qvalues}, and therefore the action $a_t$ chosen by the agent. %The parameter values used were $\alpha$ = 0.4, $\gamma =0.8$,$\epsilon = 0.3$, $\rho=0.999$ and $\xi= 0.9$. 
Algorithm \ref{alg:RL} summarizes the complete framework from Figure \ref{fig:scheme}.

In the next sections we discuss the key aspects of our approach to Problem \ref{pr:framework}. Results specifically of the external timing estimation {\color{\reviewed}are omitted here in the interest of space but} can be found in \cite{lourencco2020teaching}. %We focus on results for the internal timing mechanisms since they already include the external timing estimate, but results for this part be found in \cite{lourencco2020teaching}.% add something here about the external timing result

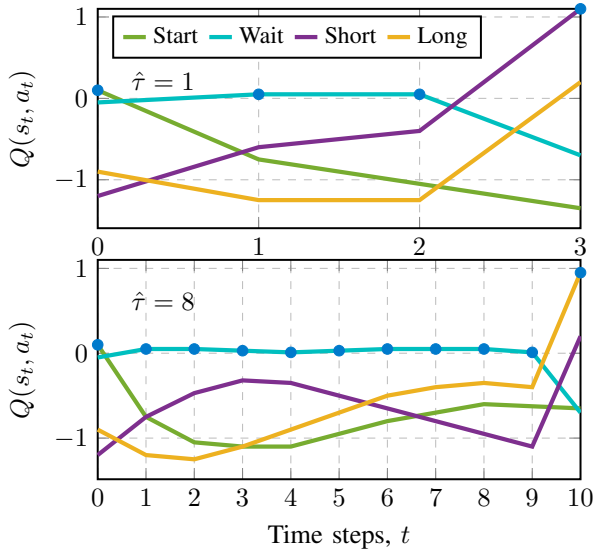
\begin{figure}[t!]
\centering
\begin{tikzpicture}
    \begin{groupplot}[
        group style={
            % set how the plots should be organized
            group size=1 by 2,
            % only show ticklabels and axis labels on the bottom
            %x descriptions at=edge bottom,
            % set the `vertical sep' to zero
            vertical sep=12pt,
        },
        % set the x axis default to logarithmic
        grid,
        height=4.5cm,
       	width=8cm,
       	ymax=1.1,
       	xmin=0,
       	ytick={-1,0,1},
        xtick ={0, 1,2, 3, 4,5,6,7,8,9,10},
        ylabel={$Q(s_t,a_t)$},
        grid style=dashed,
        %legend pos=north west,
        %legend style={draw=none, font=\footnotesize},
    ]
    % start the first plot
    \nextgroupplot[
    xmax = 3,
    legend pos=north west,
    legend columns=-1,
        legend entries={Start,Wait,Short,Long},
        legend style={/tikz/every even column/.append style={column sep=0.05cm}},
        legend style={font=\small},
        grid style=dashed,
        compat=1.11,
        legend image code/.code={
\draw[mark repeat=2,mark phase=2]
plot coordinates {
(0cm,0cm)
(0.15cm,0cm)        %% default is (0.3cm,0cm)
(0.3cm,0cm)         %% default is (0.6cm,0cm)
};%
}
    ]
     % \addlegendentry{Start}
\addplot [
    color={rgb, 1:red, 0.4660; green, 0.6740; blue, 0.1880 },
    ultra thick,  ]
    coordinates {
		(0,0.1)(1,-0.75)(2,-1.05)(3,-1.35) };
		%\addlegendentry{Start}
\addplot [
    color={rgb, 1:red, 0; green, 0.75; blue, 0.75 }, ultra thick,
    ]
    coordinates {
		(0,-0.05)(1, 0.05) (2,0.05)(3,-0.7) };
		%\addlegendentry{Wait}
\addplot [
    color={rgb, 1:red, 0.4940; green, 0.1840; blue, 0.5560}, ultra thick,
    ]
    coordinates {
		(0,-1.2)(1, -0.6)(2,-0.4)(3,1.1) };
		%\addlegendentry{Short}
\addplot [
    color={rgb, 1:red, 0.9290; green, 0.6940; blue, 0.1250 }, ultra thick,
    ]
    coordinates {
		(0,-0.9)(1,-1.25)(2,-1.25)(3,0.2)};
		%\addlegendentry{Long}
\addplot [
    blue_c,
    only marks,
    mark=*,
    ultra thin,
    ]
    coordinates {
		(0,0.1)(1, 0.05) (2,0.05)(3,1.1)};
		%\addlegendentry{$a_t$}
	\node[below right, align=center, text=black]
at (rel axis cs:0.05,0.76) {$\hat{\tau}=1 $};
		
    % start the second plot
    %
    %
    \nextgroupplot[ xlabel={Time steps, $t$}, xmax = 10,
    ]
       \addplot [
    color={rgb, 1:red, 0.4660; green, 0.6740; blue, 0.1880 },
    ultra thick, ] 
    coordinates {
		(0,0.1)(1,-0.75)(2,-1.05)(3,-1.1)(4,-1.1)(6,-0.8)(8,-0.6) (10,-0.65)};
		%\addlegendentry{Start}
\addplot [
    color={rgb, 1:red, 0; green, 0.75; blue, 0.75 }, ultra thick,
    ] 
    coordinates {
		(0,-0.05)(1,0.05)(2,0.05)(4,0.01)(6,0.05)(8,0.05)(9,0.01)(10,-0.7)};
		%\addlegendentry{Wait}
\addplot [
    color={rgb, 1:red, 0.4940; green, 0.1840; blue, 0.5560}, ultra thick,
    ]
    coordinates {
		(0,-1.2)(1,-0.75)(2,-0.47)(3,-0.32)(4,-0.35)(6,-0.65)(8,-0.95)(9,-1.1)(10,0.2) };
		%\addlegendentry{Short}
\addplot [
    color={rgb, 1:red, 0.9290; green, 0.6940; blue, 0.1250 }, ultra thick,
    ] 
    coordinates {
		(0,-0.9)(1,-1.2)(2,-1.25)(3,-1.1)(4,-0.9)(6,-0.5)(7,-0.4)(8,-0.35)(9,-0.4)(10,0.95)};
		%\addlegendentry{Long}
\addplot [
    blue_c,
    only marks,
    mark=*,
    ultra thin,
    ]
    coordinates {
		(0,0.1)(1,0.05)(2,0.05)(3,0.03)(4,0.01)(5,0.03)(6,0.05)(7,0.05)(8,0.05)(9,0.01)(10,0.95) };
	%\addlegendentry{$a_t$}
    \node[below right, align=center, text=black]
at (rel axis cs:0.05,0.9) {$\hat{\tau}=8 $};
    \end{groupplot}
\end{tikzpicture}
\caption{Evolution of the Q-values throughout two episodes with different interval durations between tones. %The maximum interval length was 8 time steps, so durations $\in [1,2,3,4]$ are Short and $\in [5,6,7,8]$ are Long. 
In the top plot, $\hat{\tau}=1$ (short interval), and in the bottom one, $\hat{\tau}=8$ time steps (long interval). {\color{\reviewed}State of the world at each time step, in the format ``(top \textbar bottom plot time step) -- State": (0 \textbar 0) -- Init, (1 \textbar 1) -- Tone, (2 \textbar 2 to 9) -- Interval, (3 \textbar 10) -- Tone.}
The chosen action at each time step is the one with the highest Q-value at that moment. {\color{\reviewed}While the initial behavior is similar}, at the end of the episode different actions have the highest Q-value, in agreement with the corresponding duration.}
\label{fig:Q-values}
\end{figure}

\subsection{Results on the Inner Core Mechanisms of the Agent}

 %The robustness of the results comes from the Monte-Carlo simulations performed.
  %The agent performs the task in Section \ref{sec:casestudy} having these two sets of microstimuli affecting the states of the experiment in each episode. 
We begin by presenting results that bring insight into the robot's inner mechanisms when using our framework, such as the TD error and the Q-values. 	The time step values on the $x$-axis correspond to the state numbers from Figure \ref{fig:RLtask}.

Figure \ref{fig:TDerror1} shows the evolution of the TD error throughout three successful episodes with the same interval duration, but each at a different training phase. The represented episodes have $\tau = 2$ time steps, which means that the second tone takes place at $t=4$ where it is followed by the reward. 
The figure shows that the TD error from \eqref{eq:delta}, \textit{i)} decreases over time as the agent learns the optimal policy when the reward is delivered; and \textit{ii)} increases when the second tone is played. %due to acting as an indication that a reward should be delivered soon.
This means that previous (conditional) stimuli teach the agent to predict reward delivery. In other words, the second tone functions as a conditional stimulus from \textit{classical conditioning} \cite{glimcher2011}. 
% These results fit observational evidence, and
%It can be seen that, as learning occurs, the TD error from \eqref{eq:delta}: \textit{i)} increases at cue onset and \textit{ii)} decreases at reward delivery. This indicates that the model learns to predict the reward from earlier stimuli, as explained in \cite{glimcher2011}. These results match empirical data, and the second tone acts as a conditional stimulus from \textit{classical conditioning}. 

Figure \ref{fig:Q-values} shows a visual representation of the evolution of the Q-values computed from \eqref{eq:Qvalues} during one episode with $\hat{\tau}=1$ (short interval), and one with $\hat{\tau}=8$ time steps (long interval), after the agent has learned the optimal policy. 
At each time step, the action with the highest Q-value is marked with a dot and is the one selected without exploration. The resulting sequence of actions is the optimal one shown in Figure~\ref{fig:RLtask} for both episodes, which means that, based on its time estimate, the agent learns to act appropriately. 
  
%Results like the ones from Figures \ref{fig:TDerror1} and \ref{fig:Q-values} can be a premonition of the success of the framework for social interactions -- rather than waiting a predetermined interval between receiving a question and answering it, it can enable agents to adjust this interval to the situation and people involved.
 
\subsection{Results on the Replication of the Behaviour}

 %(see \cite{lourenco2019}). 
The following results illustrate the behaviour of the agent in the task. 
To simplify the comparison with the performance of mice in the original experiment, we use the seconds elapsed between tones in the $x$-axis to present the results. 

 %The performance of the agent in the task is evaluated according to how far it reaches in the episode before making a wrong action.
%After the exploration phase, the agent learns to perform the correct sequence of actions until hearing the second tone. However, to receive the reward it also needs to choose the button that correctly classifies the interval duration of that episode. 

\begin{figure}[t!]
\centering 
	\begin{tikzpicture}
    \begin{axis}[
            ybar,
            grid,
            height=4.5cm,
       		width=8cm,
            symbolic x coords={0.3, 0.6, 0.9, 1.2, 1.5, 1.8, 2.1, 2.4, 2.7, 3},
            xtick ={0.6, 1.2, 1.8, 2.4, 3},
            ylabel={\# Misclassifications},
            xlabel={Interval duration, $\hat{\tau}$ [s]},
            grid style=dashed,
        ]
        \addplot [
        color=blue_c,
        fill=blue_c ,
        ]
        table[x=interval,y=carT]{\mydata};
    \end{axis}
	\end{tikzpicture} 
	\caption{Number of misclassified episodes according to the interval duration. The total number of misclassifications is 327, the average is 1.65 s and the median is 1.8 s. As for humans and animals, the intervals in the boundary between classes are the ones most commonly misclassified.}
	\label{fig:MS_8a}
\end{figure}
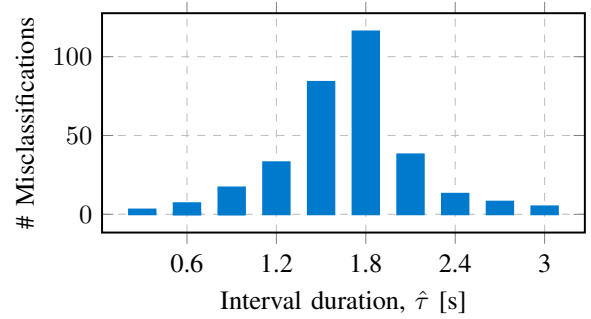

The agent learns to always acts correctly until receiving the second tone. However, even once the training phase is over, it is not always able to accurately classify the interval length of the episode since it does not always select the correct action at that moment.
Figure~\ref{fig:MS_8a} shows the number of episodes in which it misclassified the interval duration. This happens more often for intervals closer to the boundary between ``Short" and ``Long" (around $\hat{\tau}=\{1.5, 1.8\}$ seconds). % -- note that this is not the case of the episodes in Figure \ref{fig:Q-values}). 
These results show a trend also exhibited by humans and animals \cite{Gouvea2015}, which is verified for any maximum interval length.%since for humans and animals these intervals are also harder to distinguish.

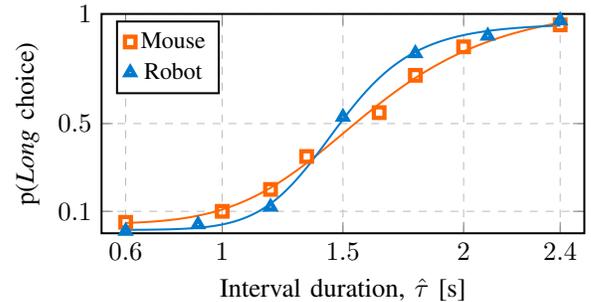
\begin{figure}[t!]
\centering 
\begin{tikzpicture}
\definecolor{clr1}{RGB}{81,82,83}
\begin{axis}[
       % axis y line=center,
       % axis x line=middle, 
       % axis on top=true,
        xmin=0.5,
        xmax=2.5,
        ymin=0,
        ymax=1,
        height=4.5cm,
        width=8cm,
        grid,
        xtick={0.6,1,1.5,2, 2.4},
        ytick={0.1,0.5,1},
        ylabel={p(\textit{Long} choice)},
        xlabel={Interval duration, $\hat{\tau}$ [s]},
        legend pos=north west,
        grid style=dashed,
        legend style={font=\small},
    ]
%\addplot [domain=-5:5, samples=50, mark=none, ultra thick, black]
 %{\FunctionF(x)};
% \node [left, blue] at (axis cs: 3.6,42) {$x^3-3x$};
 \addlegendentry{Mouse} 
  \addlegendentry{Robot}
\addplot [
    color={rgb, 255:red, 255; green, 102; blue, 0 },
    mark=square,
    ultra thick, only marks
    ]
    coordinates {
		(0.6,0.05)(1,0.1)(1.2,0.2)(1.35,0.35)(1.65,0.55)(1.8,0.72)(2,0.85)
		(2.4, 0.95) };
		
\addplot%
        [
            color={rgb, 255:red, 255; green, 102; blue, 0 },
            mark=none,
            samples=100,
            domain=0.6:2.4,
            forget plot,
        ]
        %(x,{1/(1+exp(-x))}); 
        (x, {1.0534+(0.0431-1.0534 )/(1+(x/ 1.6029)^5.6595 });  
           
\addplot [
    color=blue_c,
    mark=triangle,
    ultra thick, only marks
    ]
    coordinates {
		(0.6,0.01)(0.9,0.04)(1.2,0.12)(1.5,0.53)(1.8,0.82)(2.1,0.9)(2.4, 0.97) };
		
\addplot%
        [
            blue_c,
            mark=none,
            samples=100,
            domain=0.6:2.4,
        ]
        %(x,{1/(1+exp(-x))}); 
        (x, { 0.9574+(0.0155- 0.9574 )/(1+(x/ 1.4780)^9.2587});                
\end{axis}
\end{tikzpicture} 
\caption{Psychometric curves corresponding to the empirical probability of intervals being classified as \textit{Long}. The psychometric curve of the agent (in blue) using our framework closely matches that of the mouse (in orange). The latter was averaged over ten experiments and fitted using a logistic function.}
\label{fig:MS_8b}  

\end{figure}

Figure~\ref{fig:MS_8b} shows the empirical probability (psychometric curve) of different agents classifying different intervals as ``Long". The average performance of a mouse (in orange, from \cite{soares2016}) is qualitatively similar to that of an agent using our time perception framework (in blue). 

% xtick={0.6, 1.2, 1.8, 2.4, 3},
%0.3, 0.6, 0.9, 1.2, 1.5, 1.8, 2.1, 2.4, 2.7, 3

%Further, Figure \ref{fig:TDerror} show the decrease of the TD error as the reward starts being expected, which is in accordance with the expected [cite?].
%\begin{figure}[t]
%	\centering
%	\includegraphics[width=0.4\textwidth]{Figures/TDerror.png}
%	\caption{Evolution of the temporal-difference error, $\delta$, with the episodes.}
%	\label{fig:TDerror}
%\end{figure}

\subsection{Summary of the Results}

The above numerical experiments point to the following conclusion:

\begin{result}
%The actions of the agent in the temporal discrimination task are similar to those of the mice, showing a similar ability to classify interval durations.
Our {\color{\reviewed}simulated} robot demonstrates an ability to classify interval durations similar to that of mice, which validates the capability of our framework to use temporal information for performing time-aware actions.
\label{res:framework}
\end{result}

More specifically, Result \ref{res:framework} is supported by the similarity between the timing mechanisms of the robot and mice: % the main numerical results that show the similarity in the behavior of mice and of the robot and that allow us to claim that the robot obtained a time perception similar to that of humans and animals are the following:

\begin{itemize}
    \item The elapsed time is successfully computed from environmental data, and uncertainty in its estimation increases with the length of the interval, showing the scalar property \cite{sucala2011,lejeune2006scalar}. {\color{\reviewed}These results are presented in \cite[Figure~3]{lourencco2020teaching}};
    \item The behaviour of the TD error of our model replicates the reward prediction error of the neural dopaminergic system (it decreases with reward expectancy);
	\item The uncertainty in the classification of intervals is higher on the boundary between classes, as for humans and animals;
	\item The psychometric curves of our agent closely match the one of mice.
\end{itemize}

%%%%%%%%%%%%%%%%%%%%%%%%%%%%%%%%%%%%%%%%%%%%%%%%%%%%%%%%%%%%%%%%%%%%%%%%%%%%
\section{Estimating Timing Aspects from Behaviour }
\label{sec:estimate_framework}

%In Section \ref{sec:create_framework} we proposed a framework for proving time cognition to agents, and in Section \ref{sec:simulate_framework} we showed how it is implemented and how it works.
In the previous sections we \textit{i)} proposed a framework for replicating mechanisms of time perception and \textit{ii)} showed that it provides results consistent with biological principles, answering Problem \ref{pr:framework}.
Our next goal is to exploit this framework to gain more insight into characteristics of the brains of humans and animals, answering Problem \ref{pr:param_est}. %This is in line with one of the main goals of biology and neuroscience, which concerns understanding how humans and animals work -- from motor impulses to neural mechanisms. 
%In that sense, in this section we propose a method to estimate components of the framework used by agents as to gather insight into how their timing mechanisms work, answering Problem . 

We start by presenting characteristics that might be of interest to estimate and other works that studied them. We then detail our proposed method and conclude with numerical experiments.

\subsection{Preliminaries}
\label{ssec:parameters_prel}

In the previous sections of the paper we have demonstrated a good correspondence between real-world experiments on biological agents (i.e., mice) and that of artificial simulated agents. In other words, our mathematical framework can arguably model such behaviour well. Since there are a number of free parameters in the reinforcement learning model we could use the framework ``in reverse'' to deduce important biological characteristics of the mice. 

An example of a parameter of interest to estimate in the framework presented in Section \ref{sec:create_framework} is the number $m$ of microstimuli.
The number of microstimuli present in each agent conditions its ability to distinguish time intervals. 
This is the case since with too few microstimuli the agent has a too course perception of time and is not able to distinguish different time intervals well. This is discussed further below. 
%While with a small number of microstimuli the agent is not able to estimate between a certain number of intervals, with many it has the means to distinguish them. More details on this are given later.

Apart from the number of microstimuli, other parameters of the temporal-difference learning framework dictate the agent's behaviour and can be desired to estimate -- e.g., the learning rate $\alpha$, the discount rate $\gamma$, exploration rate $\varepsilon$, and the exploration decay $\rho$. Other parameters of the Microstimuli framework can also be estimated, such as the microstimuli decay $\xi$, and the basis functions center $\nu$ and width $\beta$.
%The parameters of reinforcement learning problems like ours that dictate the behaviour of the agent as represented in Figure \ref{fig:scheme} include the learning rate $\alpha$ and the discount rate $\gamma$. When using the $\epsilon$-greedy action selection mechanism as policy, this list is extended with the exploration rate $\varepsilon$, and the exploration decay $\rho$, and, when using the Microstimuli framework, with the number of microstimuli $m$, the decay $\xi$, and the basis functions center $\nu$ and width $\beta$.
To simplify, we focus on the estimation of one or more of these parameters, under the assumption that they are static and the others are known. 
%Nevertheless, these parameters can be independently computed using methods such as the ... [cite].

Many works have studied the question of how to infer biological parameters from the agent's behaviour \cite{eaves1978model, lillacci2010parameter, bradford2019looking}, including in reinforcement learning models, but not in a way that allows temporal information to be taken into account. %In inverse reinforcement learning \cite{ng2000algorithms} and revealed preferences \cite{mas1995microeconomic}, the behaviour of agents is used to infer characteristics of its reward and utility functions. Other methods are also used. For example, 
%In {\color{red}[cite]}, the authors use a particle filter to estimate the values of $\alpha$ and $\varepsilon$ most likely to be responsible by the behaviour of a mouse in a specific task.
%-Estimating reinforcement learning meta-parameters from animal behavioural data
   \pgfplotstableread{
   %idx g  b  p  y
	1  38 133 57 40
	%2  47 179 69 36
	3  17 121 31 38
	%4  58 178 25 39
	5  70 203 25 50
	%6  70 201 18 33
	7  83 224 16 37
	%8  86 240 26 34
    }\dataseta

   \pgfplotstableread{
   %idx g  b  p  y
 1 169 336 170 4
% 2 178 525 169 8
 3  153 600 102 52
 %4  147 726 62 81
 5  154 899 43 109
 %6  161 1109 34 127
 7  179 1406 24 153
% 8  181 1594 10 169
    }\datasetb

\pgfplotstableread{
   %idx g  b  p  y
 1 188 377 188 4
% 2 160 474 158 0
 3 189 742 185 4
% 4 171 808 155 8
 5 175 1033 4 167
% 6 174 1175 1 167
 7 181 1416 4 177
% 8 167 1464 0 162
    }\datasetc   

\pgfplotstableread{
   %idx g  b  p  y
%1 0.05409582689335394   0.37712519319938176    0.3400309119010819    0.2287480680061824
  %2 0.06716417910447761,   0.4298507462686567,   0.3074626865671642,   0.19552238805970149],
  3 0.054016620498614956   0.4487534626038781    0.2867036011080332    0.21052631578947367
 4 0.0881578947368421   0.43026315789473685  0.2355263157894737   0.24605263157894736
  5 0.11642949547218628   0.4476067270375162   0.22639068564036222   0.2095730918499353
  6 0.12928759894459102   0.5184696569920845   0.1662269129287599   0.18601583113456466
  %7 0.13216957605985039    0.47256857855361595   0.21321695760598502   0.18204488778054864
  % 8 [0.12965340179717585,   0.4672657252888318,   0.220795892169448,   0.1822849807445443]],
    }\datasetpa

	\pgfplotstableread{
   %idx g  b  p  y
 1 0.25383542538354253  0.498837749883775 0.24732682473268247 0.0
  % 2 [0.20339622641509433,  0.5969811320754717,  0.19773584905660377   0.0018867924528301887],
  3 0.17058823529411765   0.6631578947368421   0.1411764705882353   0.025077399380804954
 % [0.14448359989156953,   0.7137435619409054,   0.08891298454865817,   0.0528598536188669],
  5 0.12758540553102488    0.7476179409714153    0.033929816407157795   0.09086683709040204
  % 6 [0.11336717428087986,   0.7758037225042301,   0.013747884940778   0.09708121827411167],
  7 0.10240274599542334   0.7974828375286042   0.006864988558352402   0.09324942791762014
 % 8 [0.09328882642304989,   0.8164089950808152,   0.0014054813773717   0.08889669711876318]],
    }\datasetpb

	\pgfplotstableread{
   %idx g  b  p  y
1 0.2552204176334107   0.4988399071925754   0.24547563805104408   0.0004640371229698376
 % 2 [0.20273972602739726,   0.5992172211350294,   0.197260273972602   0.0007827788649706458],
  3 0.16914038342609772   0.6651205936920223   0.16388373531230674   0.0018552875695732839
  % 4 [0.14445034519383962,   0.7132235793945831,   0.14126394052044   0.0010621348911311736],
  5 0.12608299133606932    0.749202006383949   0.0013679890560875513   0.12334701322389421
  % 6 [0.11314308681672025,   0.7761254019292605,   0.0004019292604501   0.11032958199356913],
  7 0.10200364298724955    0.7981785063752277   0.000546448087431694   0.09927140255009108
  % 8 [0.09230001663063363,   0.8168967237651754,   0.000166306336271   0.09063695326791951]]]
    }\datasetpc

	\pgfplotstableread{
   %idx g  b  p  y
1 0.23205741626794257    0.5311004784688995    0.23205741626794257   0.004784688995215311
%  [0.19263456090651557,   0.613314447592068,   0.19263456090651557,   0.00141643059490085],
3 0.16452442159383032   0.6760925449871465   0.15809768637532134   0.0012853470437017994
%  [0.20396600566572237,    0.7053824362606232,   0.08781869688385269,   0.0028328611898017],
5 0.287012987012987  0.7077922077922078   0.0025974025974025974   0.0025974025974025974
%  [0.2812135355892649,   0.7164527421236873,   0.001166861143523206,   0.0011668611435239206],
7 0.28506271379703535  0.7092360319270239   0.004561003420752566   0.0011402508551881414
 % [0.2870249017038008,   0.709043250327654,   0.001310615989515072,   0.002621231979030144]],
}\dataseta

	\pgfplotstableread{
   %idx g  b  p  y
1 0.2522255192878338   0.49258160237388726   0.2522255192878338   0.002967359050445104
%  [0.20229885057471264,   0.5942528735632184,   0.19310344827586207,   0.010344827586206896],
3 0.17103984450923226   0.6608357628765792   0.119533527696793   0.04859086491739553
 % [0.14710743801652892,   0.7082644628099174,   0.07107438016528926,   0.07355371900826446],
5  0.12847790507364976   0.7422258592471358   0.04991816693944354   0.07937806873977087
%  [0.112375533428165,   0.7752489331436699,   0.027738264580369845,   0.08463726884779517],
7  0.10397946084724005   0.7946084724005135   0.01668806161745828   0.08472400513478819
%  [0.09592861126603458,   0.8142777467930842,   0.008365867261572783,   0.08142777467930842]],
}\datasetb

	\pgfplotstableread{
   %idx g  b  p  y
1  0.25449101796407186    0.49550898203592814   0.24850299401197604   0.0014970059880239522
 % [0.21109607577807848,   0.5899864682002707,   0.1975642760487145,   0.0013531799729364006],
3  0.17105263157894737   0.6625939849624061   0.16541353383458646   0.0009398496240601503
%  [0.14309076042518398,   0.7130008176614882,   0.1381847914963205,   0.005723630417007359],
5  0.1295754026354319   0.7445095168374817   0.0036603221083455345   0.12225475841874085
 % [0.11141120864280891,   0.7765023632680621,   0.0006752194463200541,   0.11141120864280891],
7  0.1004313000616143   0.7979051139864448   0.0012322858903265558   0.1004313000616143
 % [0.09214659685863874,   0.8162303664921466,   0.0005235602094240838,   0.09109947643979058
   }\datasetc

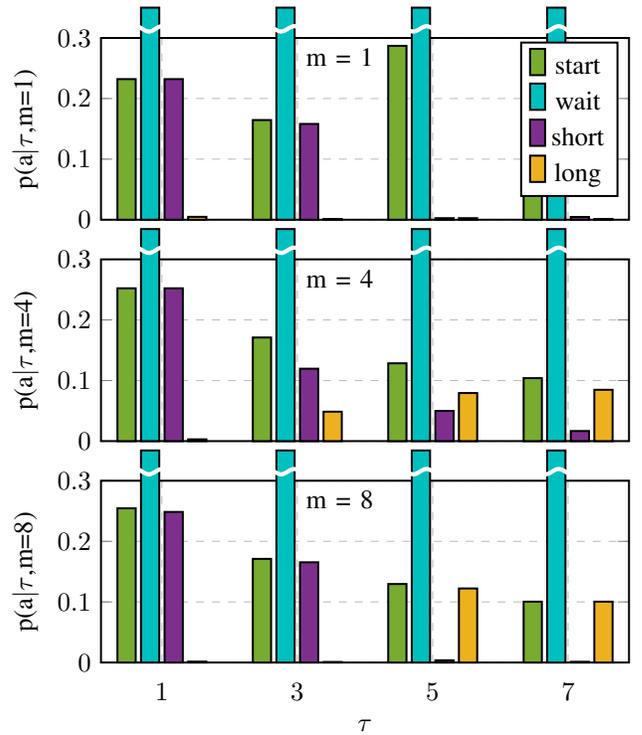
\begin{figure}[t!]
\centering
\begin{tikzpicture}
    \begin{groupplot}[
        group style={
            % set how the plots should be organized
            group size=1 by 3,
            % only show ticklabels and axis labels on the bottom
            x descriptions at=edge bottom,
            % set the `vertical sep' to zero
            vertical sep=15pt,
        },
        % set the x axis default to logarithmic
        grid,
        height=5cm,
       	width=15cm,
        %symbolic x coords={0.3, 0.6, 0.9, 1.2, 1.5, 1.8, 2.1, 2.4, 2.7, 3},
        %ylabel={Incorrect intervals},
        grid style=dashed,
        %legend pos=north west,
        %legend style={draw=none, font=\footnotesize},
    ]
    % start the first plot
    \nextgroupplot[
			ybar,
        height=4cm,
       	width=7cm,
       	x=0.9cm,
           ymin=0,
         enlarge x limits=0.15,
           ymax=0.3,        
           ylabel={p(a$|\tau$,m=1)},
           xtick=data,
          major x tick style = {opacity=0},
          minor tick length=0ex,
          bar width=0.27,
                    restrict y to domain*=0:0.35,
          %group gap/.initial=0.75cm,
          clip=false,
          after end axis/.code={ % Draw line indicating break
            \draw [ultra thick, white, decoration={snake, amplitude=1pt}, decorate] (rel axis cs:0,1.05) -- (rel axis cs:1,1.05);
        },
          legend image code/.code={
        \draw [#1] (0cm,-0.1cm) rectangle (0.2cm,0.25cm); },
            ]
  \addplot[fill={rgb, 1:red, 0.4660; green, 0.6740; blue, 0.1880 }] table[x       index=0,y index=1] \dataseta; 
   \addplot[fill={rgb, 1:red, 0; green, 0.75; blue, 0.75 }]		 table[x    index=0,y index=2] \dataseta; 
   \addplot[fill={rgb, 1:red, 0.4940; green, 0.1840; blue, 0.5560}] table[x    index=0,y index=3] \dataseta; 
   \addplot[fill={rgb, 1:red, 0.9290; green, 0.6940; blue, 0.1250 }] table[x    index=0,y index=4] \dataseta;  
    \legend{start,wait,short,long}
            \node[below right, align=center, text=black]
at (rel axis cs:0.37,1) {\text{m = 1} };
     % \addlegendentry{Start}

    % start the second plot
    \nextgroupplot[
			ybar,
        height=4cm,
       	width=7cm,
       	x=0.9cm,
           ymin=0,
         enlarge x limits=0.15,
           ymax=0.3,        
           ylabel={p(a$|\tau$,m=4)},
           xtick=data,
          major x tick style = {opacity=0},
          minor tick length=0ex,
          bar width=0.27,
                    restrict y to domain*=0:0.35,
          %group gap/.initial=0.75cm,
          clip=false,
          after end axis/.code={ % Draw line indicating break
            \draw [ultra thick, white, decoration={snake, amplitude=1pt}, decorate] (rel axis cs:0,1.05) -- (rel axis cs:1,1.05);
        },
            ]
  \addplot[fill={rgb, 1:red, 0.4660; green, 0.6740; blue, 0.1880 }] table[x       index=0,y index=1] \datasetb; 
   \addplot[fill={rgb, 1:red, 0; green, 0.75; blue, 0.75 }] table[x    index=0,y index=2] \datasetb; 
   \addplot[fill={rgb, 1:red, 0.4940; green, 0.1840; blue, 0.5560}] table[x    index=0,y index=3] \datasetb; 
   \addplot[fill={rgb, 1:red, 0.9290; green, 0.6940; blue, 0.1250 }] table[x    index=0,y index=4] \datasetb;  
            \node[below right, align=center, text=black]
at (rel axis cs:0.37,1) {\text{m = 4} };
   % start the first plot

    \nextgroupplot[
			ybar,
        height=4cm,
       	width=7cm,
           ymin=0,
           x=0.9cm,
           ymax=0.3,        
           ylabel={p(a$|\tau$,m=8)},
           xlabel={$\tau$},
           xtick=data,
          major x tick style = {opacity=0},
          minor tick length=0ex,
          bar width=0.27,
          enlarge x limits=0.15,
          restrict y to domain*=0:0.35,
          %group gap/.initial=0.75cm,
          clip=false,
          after end axis/.code={ % Draw line indicating break
            \draw [ultra thick, white, decoration={snake, amplitude=1pt}, decorate] (rel axis cs:0,1.05) -- (rel axis cs:1,1.05);
        },
            ]
  \addplot[fill={rgb, 1:red, 0.4660; green, 0.6740; blue, 0.1880 }] table[x       index=0,y index=1] \datasetc; 
   \addplot[fill={rgb, 1:red, 0; green, 0.75; blue, 0.75 }] table[x    	index=0,y index=2] \datasetc; 
   \addplot[fill={rgb, 1:red, 0.4940; green, 0.1840; blue, 0.5560}] table[x    index=0,y index=3] \datasetc; 
   \addplot[fill={rgb, 1:red, 0.9290; green, 0.6940; blue, 0.1250 }] table[x    index=0,y index=4] \datasetc;  
               \node[below right, align=center, text=black]
at (rel axis cs:0.37,1) {\text{m = 8} };
    \end{groupplot}
\end{tikzpicture}
\caption{Effect of the number of miscrostimuli $m$ on the behaviour of sets of actions $a$ of the agent for different time intervals between two stimuli, $\tau$. These statistics are the base of our model \eqref{eq:model} and were computed over one training simulation with 2000 episodes.}.
\label{fig:model}
\end{figure}

\subsection{Proposed Method for Parameter Estimation}
\label{ssec:estimate_nrMS}

%Different parameter values generate different policies, and therefore different speeds for achieving the optimal policy, or optimal behaviour. Methods to estimate these parameters are [cite] ...  . 
%In this work, we estimate the parameters by observing the agent's policy: by studying its behaviour, we infer its parameter values.

%Using a Bayesian approach we obtain:
%\begin{align}
%p(\theta|\mathcal{H}) = p(\mathcal{H}|\theta) p(\theta).
%\end{align}
%Considering uniform prior, that is, that all parameter values are equally likely, our quantity of interest is the likelihood of each parameter value, $L(\theta) = p(\mathcal{H} | \theta)$.
%This term is computed as the probability of a certain sequence of actions for different parameter values -- using a certain value, what is the probability of the agent behaving in the different ways?

To answer Problem \ref{pr:param_est}, we reformulate the question posed to: given a certain behavior, find the parameter values (broadly denoted $\theta$ here) that best explain it.
%That is, find $p(\theta|\mathcal{H})$, where the history, $\mathcal{H}$, is a set of experiments $\{(s_t,a_t,r_t)\}$ performed by the agent.

Here, the behaviour corresponds to the policy followed by the agent -- that is, the actions taken at different states. Since different parameters induce different behaviours, the most straight-forward approach would consist of analysing the correspondence between the state-action probabilities induced by the parameters and the agent's behaviour. This approach is, however, not possible for time-dependent problems since the states do not encode temporal information. The approach we propose consists of adding to this formulation the variable $\tau$ that encodes time when analysing the behaviour of the agent through the actions $a$ performed. Our model is:
\begin{equation}
p(a |\tau,\theta),
\label{eq:model}
\end{equation}
where the actions $a$ are modeled as a stochastic variable described by this probability function. It represents the distribution of actions performed with parameters $\theta$ (e.g., $\theta = [m,\alpha,\varepsilon, \dots]$) on an episode with time interval $\tau$. 

This model can be obtained from the empirical probability distribution using simulation data, and is shown in Figure \ref{fig:model} for $\theta$ being the number of microstimuli $\theta =m$. %We wish to keep this method as general as possible to be used in other problems rather than making it problem-specific, so we look at the whole action distribution and not just the final action. 
{\color{\reviewed}The statistics of episodes like the top one in Figure \ref{fig:Q-values} are shown in the sets on the left (with $\tau=1$), and those of the bottom one ($\tau=8$) are not shown but would be similar to the sets with $\tau=7$ on the right.}
The figure further shows how the ratio between certain actions changes according to the parameter (namely, $m=1,4$ and 8 microstimuli). 
For example, the ratio of ``short" and ``long" actions is only correct for $m=8$ microstimuli. 
This change can also be observed in the other variables mentioned in Section \ref{ssec:parameters_prel}, although for some more noticeably than others.

Observing the behaviour of agents in this context means then collecting information about the actions they perform at different time intervals. Thus, we collect a \textit{history} of actions
\begin{align}
\begin{aligned}
\mathcal{H}=[& \left(\tau_{1}, a_{1,1}\right), \ldots,\left(\tau_{1}, a_{1, N_{1}}\right),  \\
& \left(\tau_{2}, a_{2,1}\right), \ldots,\left(\tau_{2}, a_{2, N_{2}}\right), \ldots],
\end{aligned}
\end{align}	
where different episodes $k$ correspond to different time intervals $\tau_k$. At episode $k$, $N_k$ actions are performed, $a_{k,1}, \dots, a_{k, N_k}$.

We write the negative log-likelihood of a history $\mathcal{H}$ given parameters $\theta$ as
\begin{equation}
\begin{aligned}
L(\mathcal{H} ; \theta)=&-\sum_{k} \sum_{n=1}^{N_{k}} \log p\left(a_{k, n} \mid \tau_{k}, \theta\right),
\end{aligned}
\label{eq:likelihood}
\end{equation}
since
\begin{equation}
\begin{aligned}
p(\mathcal{H} \mid \theta) & \propto p\left(\left\{a_{k, n}\right\} \mid\left\{\tau_{k}\right\}, m\right) \\
&=\prod_{k} \prod_{n=1}^{N_{k}} p\left(a_{k, n} \mid \tau_{k}, \theta\right),
\end{aligned}
\end{equation}
assuming conditional independence among individual actions given the parameters and the corresponding episode duration.
The maximum likelihood estimator is then
\begin{equation}
\hat{\theta}=\arg \min _{\theta} L(\mathcal{H} ; \theta).
\label{eq:mlparam}
\end{equation}

This method has a limitation for imbalanced data sets, where, for distinct episodes $k$, the number of actions performed $N_k$ is very different.

Nevertheless, per the above we can estimate different parameters in the framework from empirically observing action distributions and computing the maximum likelihood estimate.

%%%%%%%%%%%%%%%%%%%%%%%%%%%%%%%%%%%%%%%%%%%%%%%%%%%%%%%%%%%%%%%%%%%%%%%%%%%%
\subsection{Numerical Experiments }

%We ran the simulations on a X-GPU laptop
We use the method described in Section \ref{ssec:estimate_nrMS} to estimate the number of microstimuli ($\theta =m$) of an agent performing the same task from Section \ref{ssec:experimental}.

\begin{figure}[t!]
\centering 
	\begin{tikzpicture}
    \begin{axis}[
            ybar,
            grid,
            height=5cm,
       		width=8cm,
            symbolic x coords={1,2,3,4,5,6,7,8},
            xtick ={1,2,3,4,5,6,7,8},
            ylabel={Normalized $L(\mathcal{H};m)$},
            xlabel={Number of micrsotimuli, $m$},
            grid style=dashed,
            ymin=0.07,
        ]
        \addplot [
        color=blue_c,
        fill=blue_c ,
        ]
        table[x=nrMS,y=likelihood]{\likelihooddata};
    \end{axis}
	\end{tikzpicture} 
	\caption{Normalized likelihood. The likelihood of each number of microstimuli increases as it approaches the true value. The maximum likelihood estimate, $\hat{m}$ \eqref{eq:mlparam} coincides with the true value, $m=4$.}
	\label{fig:likelihood}
\end{figure}
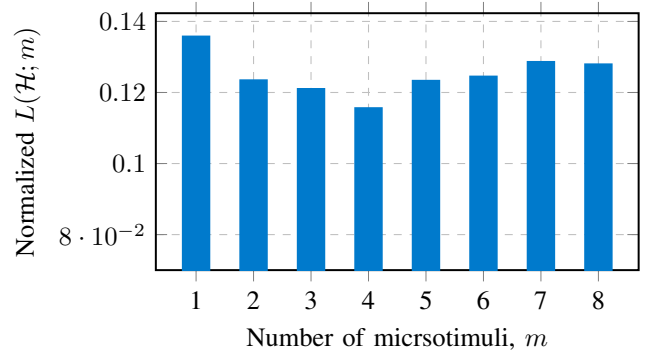

Figure \ref{fig:likelihood} shows the normalized average likelihood \eqref{eq:likelihood} computed over ten Monte Carlo simulations of 2000 episodes for the different numbers of microstimuli, when the true value is $m=4$. It can be seen that the likelihood of each number of microstimuli increases as it approaches the true value, and that the maximum likelihood estimate is correct.

We averaged the training of the model in \eqref{eq:model} over 10 train simulations of $k=2000$ episodes. When testing it in 30 test simulations, a correct maximum likelihood estimate \eqref{eq:mlparam} was obtained 100\% of the time ($\hat{m} = m$).

The above numerical experiments point to the following:
\begin{result}
Using the maximum likelihood estimator one can correctly use the behaviour of the agent to infer parameters intrinsic to the agent's internal timing mechanisms.
\label{res:parameters}
\end{result}

Since we are using a discrete model, we define \textit{accuracy} as the
percentage of times that the maximum likelihood estimate is the correct solution ($\hat{\theta} = \theta$). %For the continuous case other metrics could be used, like $|\theta-\hat{\theta}|$.
We compute the likelihood over a number of microstimuli $m$ in the set $m =\{1,\dots,8\}$ in all results presented. Let us now analyse some properties of the proposed method.\\

% [0.58, 0.26, 1.0, 1.0, 0.86, 0.44, 0.34, 0.12] accuracy for each nrMS

\pgfplotstableread[row sep=\\,col sep=&]{
    nrMS & likelihood \\
   1     & 0.13574165  \\
    2     &  0.1234268 \\ 
    3  & 0.12098436  \\
    4 & 0.11558282 \\
    5 & 0.12328322 \\
    6 & 0.12447372 \\
    7 & 0.12859388 \\
    8 &  0.12791355  \\
    }\likelihooddata

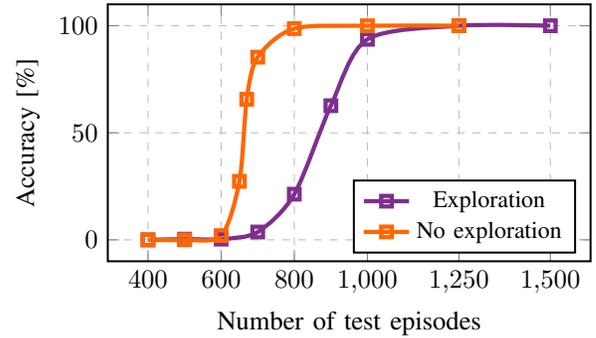
\begin{figure}[t!]
\centering 
\begin{tikzpicture}
\definecolor{clr1}{RGB}{81,82,83}
\begin{axis}[
        height=5cm,
        width=8cm,
        grid,
        xtick={400, 600, 800, 1000,1250,1500},
        xlabel={Number of test episodes},
        ylabel={Accuracy [\%]}, %Percentage of successes [\%]},
        legend pos=south east,
        grid style=dashed,
        legend style={font=\small},
    ]
    \addlegendentry{Exploration}
    \addlegendentry{No exploration}
\addplot [
    color={rgb, 1:red, 0.4940; green, 0.1840; blue, 0.5560 },
    mark=square,
    ultra thick,
    smooth,
    ]
    coordinates {
		(400,0.0)(500,0.33333333333333335)(600,0.33333333333333335)(700,3.666666666666667)(800,21.33333333)(900,62.66)(1000,93.6666)(1250, 100)(1500, 100) };
\addplot [
    color={rgb, 255:red, 255; green, 102; blue, 0 },
    mark=square,
    ultra thick,
    smooth,
    ]
    coordinates {
		(400,0.0)(500,0.0)(600,2)(650,27.33333)(670,65.66666667)(700,85.33333333)(800,98.66666667)(1000,100)(1250,100)};   
\end{axis}
\end{tikzpicture} 
\caption{Evolution of the percentage of successes (accuracy) of the estimator as the number of test episodes increases. After around 1000 test episodes, the original estimator (that admits exploratory actions, in purple) has a 100\% accuracy while the estimator that ignores exploratory actions (in orange) has the same accuracy after only 800 episodes.}
\label{fig:nEpisodes}  
\end{figure}

\subsubsection{Number of testing episodes}

Figure \ref{fig:nEpisodes} shows the importance of the number of test episodes for the accuracy of the estimator.
More specifically, the purple line shows how the accuracy (the percentage of times that $\hat{m} = m$) increases (averaged over 300 test simulations to improve the precision) as the number of test episodes increases.
This result is related to the level of exploration of the agent studied next, since the reason for the low accuracy in low numbers of test episodes is the exploratory (i.e., random) behaviour of the agent before learning the optimal policy.\\

\subsubsection{Exploration}

%Examples of parameters of the agents that do not need to be known to use this method are the exploration rate $\varepsilon$ and decay $\rho$ of its exploration algorithm. However, a
Since our model uses the statistics of all actions performed throughout all episodes, its accuracy is affected by the agent following the optimal policy or performing exploratory actions, as previously shown. %Accounting for exploratory actions in our model makes it therefore less accurate. 
If the exploration parameters $\varepsilon$ and $\rho$ of the agent are known (or estimated using the method in Section \ref{ssec:estimate_nrMS} with $\theta=\varepsilon$, or directly from the agent's behaviour), we can train our model only with knowledge-exploiting behaviours. This corresponds to only observing the actions performed for small values of $\varepsilon$ (e.g., $\varepsilon < 0.01$).

The orange line in Figure \ref{fig:nEpisodes} shows that, as expected, observing only knowledge-exploiting actions is advantageous for the accuracy of the estimator, increasing for a comparative number of testing episodes.\\ %(By around 5\%). Still with an average of 10 train and 300 test simulations.\\

%Figure \ref{fig:epsilon} shows the influence of the exploration parameters in the accuracy of the estimator. 

\subsubsection{Sensitivity} 

The parameters inherent to the agents (that we assume static and known) can be estimated but are typically not known with certainty. To explore the sensitivity of our estimator to the uncertainty in the parameters, we perturb them when estimating the number of microstimuli.

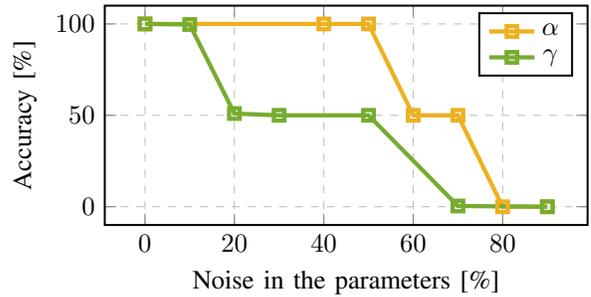
\begin{figure}[t!]
\centering 
\begin{tikzpicture}
\definecolor{clr1}{RGB}{81,82,83}
\begin{axis}[
        height=4.5cm,
        width=8cm,
        grid,
        xlabel={Noise in the parameters [\%]},
        ylabel={Accuracy [\%]}, %Percentage of successes [\%]},
        legend pos=north west,
        grid style=dashed,
        legend pos=north east,
    ]
    \addlegendentry{$\alpha$}
    \addlegendentry{$\gamma$}
\addplot [
    color={rgb, 1:red, 0.9290; green, 0.6940; blue, 0.1250 },
    mark=square,
    ultra thick,
    %smooth,
    ]
    coordinates {
		(0,100)(40,100)(50,100)(60,50)(70,50)(80,0)(90,0)};
		%(20,100)(40,60)(60,50)(80,10.500000000000001)};
		% alpha, with 50 test experiments e 1 training: 
		% (10,100)(30,90.8)(50,60.8)(70,50)
\addplot [
    color={rgb, 1:red, 0.4660; green, 0.6740; blue, 0.1880 },
    mark=square,
    ultra thick,
    %smooth,
    %dashed,
    ]
    coordinates {
		(0,100)(10,99.6666666)(20,51)(30,50)(50,50)(70,0.4)(90,0)};
		%(20,109)(40, 80)(60,40)(80,60) };
\end{axis}

\end{tikzpicture} 
\caption{Evolution of the accuracy of the estimator as the uncertainty (noise) in the parameters $\alpha$ and $\gamma$ increases.}
\label{fig:sensitivity}  
\end{figure}

Let us choose $\alpha$ and $\gamma$ as our parameters with uncertainty. 
We perturb the true values with noise when training the model (over 5 train simulations), and compute the average accuracy between the noisy values (e.g., $\alpha \; \pm$ noise) over 5 test simulations. Figure \ref{fig:sensitivity} shows the average accuracy for percentages of noise between 0\% and 80\% in the parameters.
The large discrete steps seen on the accuracy are due to the average being made between two very different values of the parameter ($\pm$ noise). To obtain a smoother curve, the noise could instead be sampled from a uniform or Gaussian distribution.  
%We add random zero-mean Gaussian perturbations of a certain standard deviation to $\alpha$ when training the model. The average percentage of successes over 10 chosen $\alpha$ with this uncertainty gives us an average success percentage for this uncertainty. The average percentage of successes over 10 chosen $\alpha$ with this uncertainty gives us an average success percentage for this uncertainty. 
%In Figure \ref{fig:sensitivity} test the effect that standard deviations between 0\% and 50\% have on the successes of the estimator.

We conclude that the method is robust to small perturbations in both of the parameters studied, and a similar analysis can be done for other parameters of interest.\\

\subsubsection{Scalability} 

Analysing the behaviour of the agent for many different parameter values and different time intervals can be computationally heavy and impractical to implement in real life experiments. This problem is aggravated for large action spaces.
In this section, we evaluate the scalability of the estimator by exemplifying its behaviour for situations where only specific actions or certain time intervals between tones are observed. 
%This is not the best method in the world because we would need to compute all the statistics for all the possible values of each parameter. 
%
For example, it can either be \textit{i)} a whole action (e.g., the ``wait" action is not counted), or \textit{ii)} a whole time interval (e.g., intervals of $\tau=3 s$ are not tested), or even \textit{iii)} random combinations of both, that are not observed.
Whichever the situation, some intervals and actions have a higher impact for the accuracy of the estimation than others. For example, for this particular task, the ``short" and ``long" actions present more information about the knowledge of the agent than others. As can be seen in Figure \ref{fig:model}, their probability distributions change significantly according to the number of microstimuli.

%In practice, this corresponds to removing certain data from the history $\mathcal{H}$ and summing over less $k$ and/or $n$ in \eqref{eq:likelihood}.
Table \ref{tab:scalability} shows the accuracy of the estimator computed when different sets of parameters are removed over 50 test simulations. For example, the second row corresponds to not observing the actions of the agent %number of times the agent presses the ``long" button nor testing its behaviour 
in time intervals between 2 and 8 time steps. An accurate estimate of the number of microstimuli is still obtained 78\% of the time, showing that the method is flexible with the testing situations.
%\subsubsection{Distance Measure }
%How does it change with different distance measures? E.g. Euclidean etc
%We use the KL divergence as measure of comparison between distributions, because ... However, other measures ...\\

\begin{table}[t!]
\centering
\normalsize
\caption{Variation of the accuracy with the unobserved parameters. }
\begin{tabular}{ |c|@{\hskip 0.23in}c@{\hskip 0.23in}||c|  }
 \hline
 \multicolumn{2}{|c||}{Unobserved parameters } & Accuracy\\\cline{1-2}
% \hline
  $a$ & $\tau$& \\
 \hline
 -   & - & 100 \% \\
% -   &  5 - 8 & 100 \% \\
 -   & 2 to 8 & 78 \% \\
 Long   & 2 to 8 & 40 \% \\
  Long   & - & 100 \% \\
 Wait, Short, Long       & - &  2 \% \\
 Wait, Short   & 4 & 0 \% \\
 \hline
\end{tabular}
\label{tab:scalability}
\end{table}

\section{Conclusion}
\label{sec:conclusions}
This paper investigated characteristics of neural mechanisms involved in time perception, taking into consideration multiple aspects of these timing mechanisms to design an end-to-end decision-making framework, whose parameters can be inferred from the behaviour of an agent.

\subsection{Summary}
The first main focus of this work was to create a time perception framework capable of producing time-aware actions (Section \ref{sec:create_framework}). This framework consists of a combination of two known timing sources: internal neuronal mechanisms and external stimuli. For the former we replicate dopaminergic behaviour by means of a temporal-difference learning algorithm with a feature representation called \textit{Microstimuli}. For the later, we estimated of the passage of time from environmental data by exploiting results from Gaussian processes.

We validated the framework in a simulated robot, in Section \ref{sec:simulate_framework}, and compared its behaviour to that of real-world mice. The ability of an agent using the proposed algorithm to perceive time and succeed in time-dependent tasks was validated in numerical results by the comparison with the behaviour of real animals (see Result \ref{res:framework}). We presented coherent results in both sources of time perception: both in its instrinsic mechansims of interpreting time, as well as in its performed actions. We concluded the former by identifying features known to be present in humans and animals, and the latter by comparing its actions with those of mice in the same task.

The second main focus of this work was a method presented in Section \ref{sec:estimate_framework} to infer biological insight about the framework used by agents for time perception. We used a maximum likelihood estimator to deduce biological characteristics of the timing functions of agents by estimating parameters they use for perceiving time. In particular, we showed that we are able to estimate the number of microstimuli of an agent given its empirical action probability distribution, using the maximum likelihood estimator (see Result \ref{res:parameters}). We further showed \textit{i)} how the estimate improves with the number of testing episodes, \textit{ ii)} how knowing the exploration rate of the agent improves the estimation process, \textit{iii)} that uncertainty in the other parameters does not affect greatly the estimate, and \textit{iv)} that the method is scalable and works with few testing observations.
% We estimated these parameters based on their behaviour, by modeling the empirical probability distribution of its actions.

\subsection{Future Work}

There are several interesting extensions that we would like to investigate in future work. First, implementing the framework on a real robot and having it perform the same real-world task as mice would give valuable experimental data that can be used to validate and modify our proposed framework. 
Since the mechanisms controlling temporal judgments are believed to vary across time scales, we would like to study the behaviour of our framework outside the time scale of seconds and study how it can be adapted.
{\color{\reviewed}With small changes, our framework could be applied to retrospective timing or duration reproduction tasks. However, being the goal of our work to replicate biological mechanisms, further studies should be conducted in order to correctly incorporate specific underlying neural mechanisms.}
%We would also like to study how the framework behaves for other time scales and how it can be adapted, since dopaminergic neurons are believed to only control temporal judgments on a time scale of seconds.

Second, it would be highly interesting to estimate biological characteristics of various animals (and humans) by using the algorithms discussed in Section \ref{sec:estimate_framework}.
By observing their actions we could estimate their intrinsic timing parameters, and validate them by comparing them with the actions of simulated agent with those parameters. %In that case we should also consider the estimation of dynamic parameters. %Further, our method can be extended for estimating the parameters of interest when the others are not known, by marginalizing their likelihood. 
%However, the method has the limitations of not being able to handle continuous nor dynamic parameters. %An alternative to empirically compute the action distribution would be to include it in a cost function an use a simple solver such as Bayesian optimization. 
%
Another relevant question concerns how knowledge about the animal's inherent parameters can be obtained by observing its behaviour in different tasks. 

Finally, we would also like to investigate how more sophisticated methods for inverse learning problems can be combined with the timing framework discussed in this paper. For example, methods from inverse reinforcement learning and revealed preferences (e.g., \cite{ng2000algorithms, mas1995microeconomic}) or inverse filtering (e.g., \cite{mattila_inverse_2017, mattila_estimating_2019}) could be used to infer the internal beliefs of the agents based on their behaviour.

%use deep learning to analyse the non-parametric distribution of the environmental data instead of parametrized Gaussian processes.
% This paves the way for more advanced learning, such as in interval reproduction tasks performed by humans, or in learning the parameters of the neural network they use, in deep learning.\\

% or


\appendices
\section{External Timing}
\label{sec:appendix_external}

In \cite{BI}, the authors proposed a method to estimate the elapsed time $\tau$ from environmental data $\mathcal{O}$ using a Bayesian framework. Under uniform prior, the maximum likelihood estimate of the elapsed time is given by the maximum of $p(\mathcal{O}|\tau)$ \cite{ljung1987} and corresponds to the probability of observing $\mathcal{O} = \{y_t(i)\}_{i=1}^{M}$ during the interval $\tau$. This probability can be modeled as a zero-mean joint Gaussian distribution over the $N$ observations of all $M$ independent sensors \cite{gaussian2007}. With $t_1 = 0$ and $t_N = \tau$, this is given by %refers to $y_0(i), \dots ,y_{\tau}(i)$, for $i=1,\dots,M$: % 
\begin{equation}
p(y_{t}(i)|\tau) = \mathcal{N}(y_t(i); 0, K_\Omega) = \dfrac{e^{-\frac{1}{2}y_{t}(i)K^{-1}_\Omega y_{t}^T(i)}}{\sqrt{\det(2\pi K_\Omega)}} .
\label{eq:equationBI} 
\end{equation}
This joint distribution includes a kernel function $K_\Omega$ that is parametrized by $\Omega$ and expresses the variation of the process between time steps. It has been observed that the power spectrum of the observations (the rhythm of change of the natural environment) resembles that of the Ornstein-Uhlenbeck (OU) function \cite{OUcov1930}:
\begin{equation}
K_{\lambda,\sigma}(\tau) = e^{-\lambda |\tau|} + \sigma^2 \psi(\tau).
\label{eq:OU}
\end{equation}
Here, $\psi(0) =1$ and $\psi(\tau) =0$ for $\tau \neq 0$, and $\Omega =[\lambda, \sigma]$ are the hyperparameters of the model. We estimate them using Bayesian model selection, by maximizing the logarithm of the likelihood with respect to $\Omega$ \cite{ljung1987}. This involves computing the respective derivatives:
\begin{equation}
\frac{\partial}{\partial\Omega_j} \log p(y_t(i)|\Omega) = -\frac{1}{2} tr(\phi\phi^T - K^{-1}_\Omega ) \frac{\partial K_\Omega}{\partial\Omega_j},
\label{eq:Gderivates}
\end{equation}
where $\phi = K^{-1} y_t(i)$.

Hence, the external timing problem is solved by identifying the values for $\lambda$ and $\sigma$ that make the properties of the Gaussian process most similarly approximate the ones of the environmental data. The robot's estimate of the elapsed time between the two stimuli, $\hat{\tau}$, is thus obtained by computing the maximum likelihood estimate of \eqref{eq:equationBI}.
%\section{}

%\section*{Acknowledgment}

\ifCLASSOPTIONcaptionsoff
  \newpage
\fi



%\bibliographystyle{IEEEtran}

%\begin{thebibliography}{1}


%\end{thebibliography}
\bibliographystyle{ieeetr}
\bibliography{L4DCRef.bib}


\vspace{5cm}

\begin{IEEEbiography}[{\includegraphics[width=1in,height=1.25in,clip,keepaspectratio]{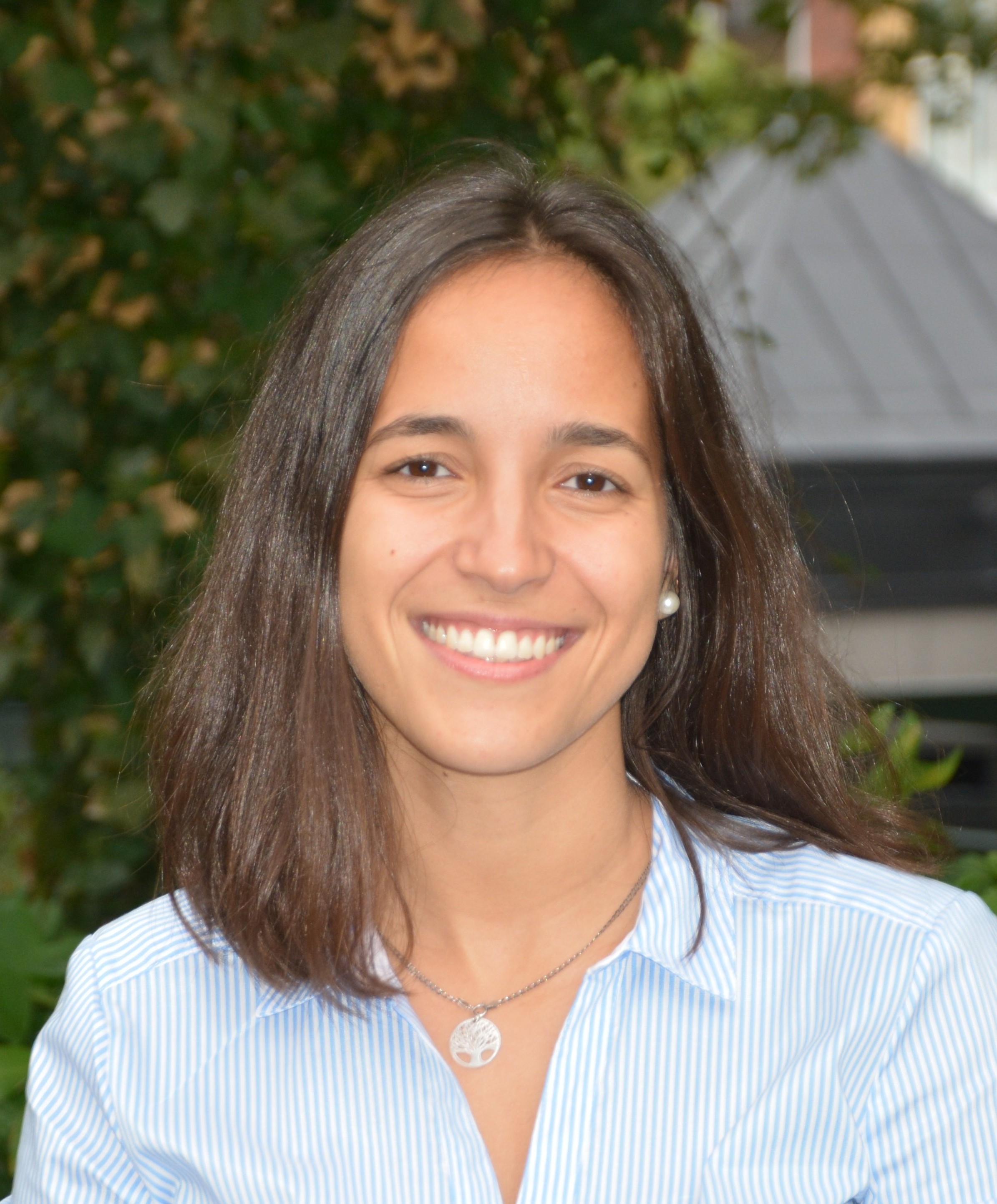}}]{In\^es Louren\c{c}o} (Student Member, IEEE) graduated from the Electrical Engineering and Computer Science programme of Instituto Superior T\'{e}cnico, Portugal, in 2018 with a M.Sc. in decision systems and control. During that time she was an intern at the Gulbenkian Science Institute and at INESC-ID, and collaborated with the Champalimaud Foundation. After receiving the KTH Electrical Engineering Scholarship of Excellence she started her Ph.D. studies at KTH Royal Institute of Science, in Stockholm, Sweden. Her primary research interests are within control of stochastic dynamical systems and biologically-inspired robotics.
\end{IEEEbiography}

\begin{IEEEbiography}[{\includegraphics[width=1in,height=1.25in,clip,keepaspectratio]{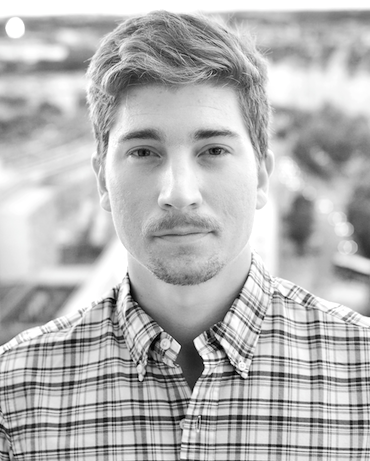}}]{Robert Mattila} (Student Member, IEEE) graduated from the Engineering Physics programme of KTH Royal Institute of Technology, Stockholm, Sweden, in 2015 with an M.Sc. in systems, control and robotics. The same year, he was awarded the KTH Electrical Engineering Scholarship of Excellence. In 2020, he received his Ph.D. degree from KTH by submitting the thesis \emph{Hidden Markov Models: Identification, Inverse Filtering and Applications}. He has been a Visiting Researcher at the California Institute of Technology (Caltech), USA, the University of British Colombia (UBC), Canada, and Cornell University, USA. His primary research interests are within inference and control of stochastic
dynamical systems.
\end{IEEEbiography}

\begin{IEEEbiography}
[{\includegraphics[width=1in,height=1.25in,clip,keepaspectratio]{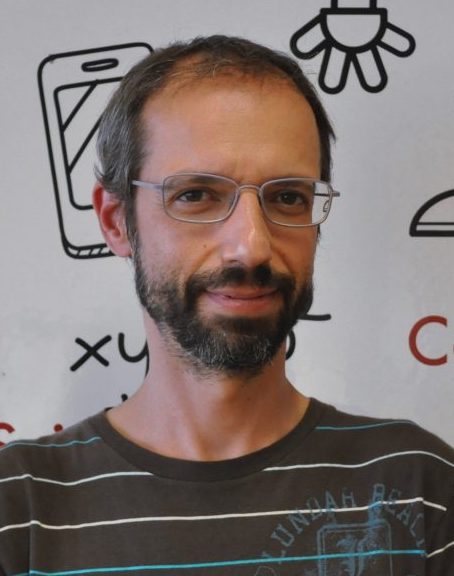}}]{Rodrigo Ventura} (Assistant Professor) received the Licenciatura (1996), M.Sc. (2000), and PhD degree (2008), in Electrical and Computer Engineering from Instituto Superior Técnico (IST), Lisbon, Portugal. He is a (tenured) Assistant Professor at IST, and a permanent member of Institute for Systems and Robotics (ISR-Lisboa). He has published more than 130 publications in peer-reviewed international journals and conferences, on various topics intersecting Robotics and Artificial Intelligence. He is also co-inventor of several national and international patents on innovative solutions for robotic systems. He is a founding member of the Biologically-Inspired Cognitive Architecture society and alumni of the International Space University. He participated in several international and national research projects. Broadly, his research is focused on the intersection of Artificial Intelligence and Robotics, with particular interests in Human-robot collaboration, Mobile manipulation, Robot planning and control, Cognitive robotics, and Biologically inspired cognitive architectures. This research is mostly driven by research questions elicited by application in space robotics, aerial robots, social service robots, and urban search and rescue robots.
\end{IEEEbiography}

%\newpage

\begin{IEEEbiography}[{\includegraphics[width=1in,height=1.25in,clip,keepaspectratio]{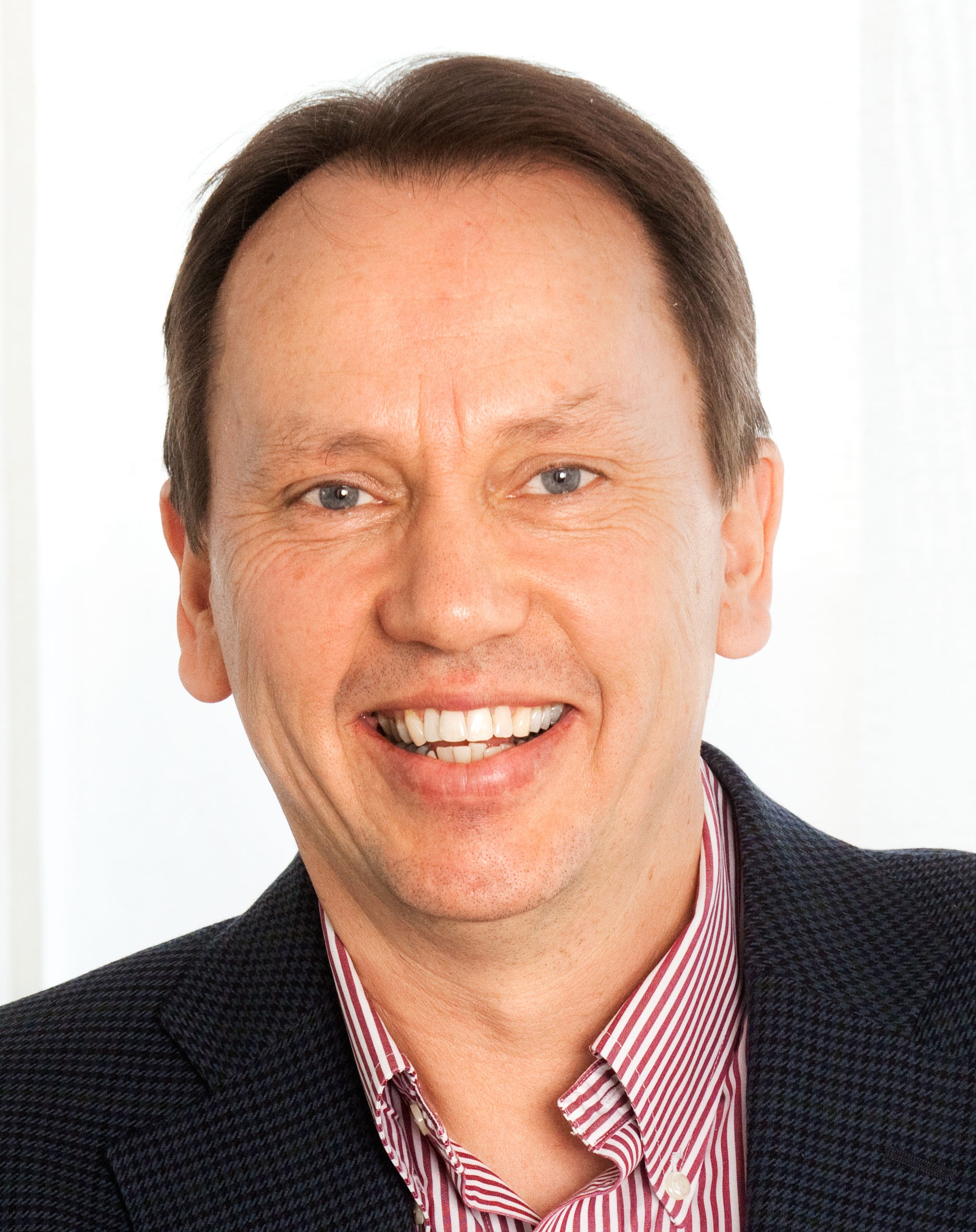}}]{Bo Wahlberg} (Fellow IEEE) received his M.Sc. degree in Electrical Engineering in 1983 and his Ph.D. degree in 1987 from Link\"{o}ping University, Sweden. In December 1991, he became Professor of the Chair of Automatic Control at KTH Royal Institute of Technology, Stockholm, Sweden He is a Fellow of the IEEE and IFAC for his contributions to system identification.  Dr.  Wahlberg was a co-founder of Centre of Autonomous Systems and the Linnaeus Center ACCESS on networked systems at KTH. He is a member of the program management group of the Wallenberg AI, Autonomous Systems and Software Program (WASP). Dr.  Wahlberg has received several awards, including the IEEE Transactions on Automation Science and Engineering Best New Application Paper Award in 2016. He was a plenary speaker at the 35th Chinese Control Conference in 2016. His research interests include system identification, modeling and control of industrial processes, machine learning and statistical signal processing with applications in automated transportation systems.
\end{IEEEbiography}


%\vfill


\end{document}